\documentclass[a4paper,11pt]{article}

\usepackage{amsmath,amssymb,bbm}
\usepackage{ascmac}
\newdimen\tbaselineshift 
\usepackage{cancel}
\usepackage{xcolor} 
\usepackage{graphicx}
\usepackage{fancybox}
\usepackage{shadow}
\usepackage{multirow}
\usepackage{upgreek}
\usepackage[bookmarks=true,bookmarksnumbered=true,bookmarkstype=toc]{hyperref}
\hypersetup{
pdfauthor={Tetsuji Kimura},
colorlinks={true}, 
linkcolor={blue}, 
urlcolor={blue}, 
filecolor={brown}, 
citecolor={blue}
}
\usepackage{marvosym,wasysym}
\usepackage{pifont}
\usepackage{arydshln}
\usepackage{fancyheadings} 
\usepackage[all]{xy}
\usepackage{ulem}
\makeatletter

 \@addtoreset{equation}{section}
\makeatother

\newcounter{Enumerate}

\DeclareFontFamily{U}{rsf}{}
\DeclareFontShape{U}{rsf}{m}{n}{
  <5> <6> rsfs5 <7> <8> <9> rsfs7 <10-> rsfs10}{}
\DeclareMathAlphabet\Scr{U}{rsf}{m}{n}

\usepackage[mathscr]{eucal}

\newcommand{\del}{\partial}

\newcommand{\half}{\frac{1}{2}}

\newcommand{\ls}{\ \ \ \ \ }
\newcommand{\wt}{\widetilde}

\newcommand{\ve}{\varepsilon}
\newcommand{\ol}{\overline}

\newcommand{\bsubeq}{\begin{subequations}}
\newcommand{\esubeq}{\end{subequations}}

\newcommand{\noi}{\noindent}

\newcommand{\nn}{\nonumber}

\newcommand{\I}{{\rm i}}

\newcommand{\N}{\mathcal{N}}

\renewcommand{\d}{{\rm d}}
\newcommand{\e}{{\rm e}}

\newcommand{\slb}{\scalebox}

\def\+{{+\!\!\!+}} 

\begin{document}
\allowdisplaybreaks{

\thispagestyle{empty}


\begin{flushright}
TIT/HEP-643 
\end{flushright}

\vspace{45mm}

\noi
\slb{2.3}{$\N=(4,4)$ Gauged Linear Sigma Models}

\vspace{5mm}

\noi
\slb{2.3}{for Defect Five-branes}

\vspace{15mm}

\noi
{\renewcommand{\arraystretch}{1.7}
\begin{tabular}{cl}
\multicolumn{2}{l}{\slb{1.2}{Tetsuji {\sc Kimura}}}
\\
& {\renewcommand{\arraystretch}{1.0}
\begin{tabular}{l}
{\sl
Department of Physics,
Tokyo Institute of Technology} 
\\
{\sl Tokyo 152-8551, JAPAN}
\\
\slb{0.9}{\tt tetsuji \_at\_ th.phys.titech.ac.jp}
\end{tabular}
}
\end{tabular}
}

\vspace{20mm}


\slb{1.1}{\sc Abstract}
\begin{center}
\slb{.95}{
\begin{minipage}{.95\textwidth}
\parindent=6mm%
We study two-dimensional $\N=(4,4)$ gauged linear sigma model (GLSM). 
Its low energy effective theory is a nonlinear sigma model whose target space gives rise to a configuration of five-branes in string theory.
In this article we focus on sigma models for NS5-branes, KK5-branes and an exotic $5^2_2$-brane.
In particular, we carefully analyze the GLSM for an exotic $5^2_2$-brane whose background configuration is multi-valued.
The exotic $5^2_2$-brane is a concrete example of nongeometric configuration in string theory.
We find that the exotic feature originates from the string winding coordinate in a very clear way. 
In order to complete this analysis, 
we propose a duality transformation formula which converts an $\N=(2,2)$ chiral superfield in F-term to a twisted chiral superfield coupled to an unconstrained complex superfield.

This article is a short review based on \cite{Kimura:2013fda} in collaboration with Shin Sasaki.
\end{minipage}
}
\end{center}

\newpage
\section{Introduction}
\label{sect:introduction}

String theory contains not only strings but also various extended objects such as D-branes \cite{Polchinski:1995mt} and NS5-branes \cite{Strominger:1990et}. 
They are strongly related to each other via string dualities.
We focus on five-branes, for instance.
A D5-brane becomes an NS5-brane via S-duality.
An NS5-brane is dualized to an Kaluza-Klein (KK) monopole (or referred to as an KK5-brane) \cite{Sorkin:1983ns} if T-duality is performed along one transverse direction of the NS5-brane.
Branes of codimension three or more are called {\it standard} branes.

Performing string dualities in lower spacetime dimensions, 
one often encounters branes of codimension two or less.
They are called {\it exotic} branes \cite{Blau:1997du, Obers:1998fb}.
Nowadays, in particular, branes of codimension two are called {\it defect} branes \cite{Bergshoeff:2011se}.
They are exotic because of two reasons in the supergravity framework (see appendix \ref{app:SUGRA}):
(1) the IR divergence in their background configurations, and 
(2) non-trivial monodromy structures originated from string duality.
We expect that such exotic features would reveal aspects of quantum gravity in string theory.

In order to investigate exotic branes in string theory,
we introduce two-dimensional supersymmetric gauge theory coupled to certain matter multiplets.
This is called gauged linear sigma model (GLSM) \cite{Witten:1993yc}.
GLSM is a very powerful model because it can be regarded as the UV completion of nonlinear sigma model (NLSM) as string worldsheet theory.
Originally $\N=(2,2)$ GLSM has been widely utilized in string theory compactified on Calabi-Yau space and its mirror symmetry \cite{Hori:2000kt}.
$\N=(4,4)$ GLSM, obtained by quiver gauge theory \cite{Douglas:1996sw, Johnson:1996py} and brane configuration \cite{Giveon:1998sr}, 
is interpreted as an effective theory on D1-branes or compactified D2-branes intersecting with NS5-branes.
An $\N=(4,4)$ GLSM with $k$ charged hypermultiplets with one neutral hypermultiplet describes the background configuration of $k$ NS5-branes or $k$ KK5-branes \cite{Tong:2002rq,Harvey:2005ab, Okuyama:2005gx}.
Since an exotic $5^2_2$-brane \cite{deBoer:2010ud, Kikuchi:2012za, deBoer:2012ma} is found by T-duality from an KK5-brane, 
we should be able to investigate the exotic feature of the $5^2_2$-brane in the $\N = (4,4)$ GLSM framework.
This is the main motivation of our work \cite{Kimura:2013fda}.

The $\N=(4,4)$ GLSM for NS5-branes or for KK5-branes is represented in terms of $\N=(2,2)$ supermultiplets under $SU(2)$ R-symmetry.
T-duality on the target space configuration of its low energy effective theory can be traced as the Legendre transformation in the GLSM level.
Indeed the duality transformation from the GLSM for an NS5-brane to that for a KK5-brane can be completed by the Ro\v{c}ek-Verlinde formula \cite{Rocek:1991ps, Hori:2000kt}.
However, there are technical difficulties when we construct the GLSM for an exotic $5^2_2$-brane from the GLSM for a KK5-brane via T-duality.
Because, in this case, the duality transformation associated with T-duality should handle an $\N=(2,2)$ chiral superfield in F-term.
Roughly speaking, F-term dictates interaction terms {\it without} derivatives.
This implies that there are no (global) shift symmetry in two dimensions, and no isometry on its target space.
Even in this situation, we can find a trick to justify that the duality transformation of the chiral superfield in F-term generates the correct T-duality on the target space \cite{Kimura:2013fda, Kimura:2014bxa, Kimura:2014aja}.
By virtue of this trick, we can investigate the UV completion of the string sigma model whose target space is the background configuration of the exotic $5^2_2$-brane \cite{Kimura:2013fda, Kimura:2013zva, Kimura:2013khz}.
We will understand that the exotic feature of the $5^2_2$-brane originates from the string winding coordinate.
The contribution of the winding coordinate is nothing but an evidence of stringy effect to the background configuration beyond supergravity.


The structure of this article is as follows.
In section \ref{sect:N44-gauge} we consider two-dimensional $\N=(4,4)$ supersymmetric gauge theories.
First we prepare $\N=(2,2)$ Lagrangians in terms of $\N=(2,2)$ superfields.
Second we impose $SU(2)^3$ R-symmetry which should be involved in $\N=(4,4)$ system.
In section \ref{sect:T-dual-F} we discuss duality transformation formulae of $\N=(2,2)$ superfields. In particular, the duality transformation of chiral superfields in D-terms and F-terms will play a significant role in dualities among $\N=(4,4)$ theories.
In section \ref{sect:defect5} we study GLSMs and their IR limit. 
It turns out that the target spaces of the IR theories denote the background configurations of defect five-branes.
In particular, we carefully analyze the GLSM for an exotic $5^2_2$-brane and derive the exotic structure.
Section \ref{sect:summary} is devoted to summary and discussions.
In appendix \ref{app:conventions} we write down the conventions in this article.
In appendix \ref{app:SUGRA} we gather supergravity solutions of five-branes.


\section{2D $\N=(4,4)$ gauge theory}
\label{sect:N44-gauge}

In this section we construct an $\N=(4,4)$ gauge theory in terms of $\N=(2,2)$ supermultiplets under $SU(2)^3$ R-symmetry.
In this article we focus only on abelian gauge symmetry.

\subsection{Gauge multiplets}

An $\N=(4,4)$ gauge multiplet involves a vector field $A_m$, four Weyl fermions $(\lambda_{\pm}, \wt{\lambda}_{\pm})$, two complex scalars $(\sigma, \phi)$, one real auxiliary scalar $D_V$ and one complex auxiliary scalar $D_{\Phi}$.
The subscripts of the Weyl fermions represent their chirality.
All of them take values in the adjoint representation of a gauge group.
It is convenient to express them in terms of $\N=(2,2)$ superfields $V$ and $\Phi$.
The former is a vector superfield and the latter is an adjoint chiral superfield.
Their explicit forms are 
\bsubeq \label{SF-GM}
\begin{align}
V \ &= \
- \theta^+ \ol{\theta}{}^+ (A_0 + A_1)
- \theta^- \ol{\theta}{}^- (A_0 - A_1)
- \sqrt{2} \, \theta^- \ol{\theta}{}^+ \sigma
- \sqrt{2} \, \theta^+ \ol{\theta}{}^- \ol{\sigma}
\nn \\
\ & \ \ \ \ 
- 2 \I \, \theta^+ \theta^- (\ol{\theta}{}^+ \ol{\lambda}{}_+ + \ol{\theta}{}^- \ol{\lambda}{}_-)
+ 2 \I \, \ol{\theta}{}^+ \ol{\theta}{}^- (\theta^+ \lambda_+ + \theta^- \lambda_-)
+ 2 \, \theta^+ \theta^- \ol{\theta}{}^+ \ol{\theta}{}^- D_V
\, , \label{V} \\
\Phi \ &= \ 
\phi 
+ \I \sqrt{2} \, \theta^+ \wt{\lambda}{}_+ 
+ \I \sqrt{2} \, \theta^- \wt{\lambda}{}_-
+ 2 \I \, \theta^+ \theta^- D_{\Phi}
+ \ldots
\, . \label{Phi}
\end{align}
\esubeq
Here we take the Wess-Zumino gauge.
Since we adopt the Lorentz signature of two-dimensional spacetime,
the fermions $\ol{\lambda}{}_{\pm}$ and $\ol{\wt{\lambda}}{}_{\pm}$ are hermitian conjugate $\ol{\lambda}{}_{\pm} = (\lambda_{\pm})^{\dagger}$ and $\ol{\wt{\lambda}}{}_{\pm} = (\wt{\lambda}_{\pm})^{\dagger}$. 
The symbol ``$\ldots$'' in $\Phi$ implies derivative terms governed by the covariant derivatives $D_{\pm}$ and $\ol{D}{}_{\pm}$ defined in (\ref{D_olD}).
The $\N=(2,2)$ vector multiplet $\{ A_m, \sigma, \lambda, D_V \}$ is often described in terms of a twisted chiral superfield $\Sigma$ defined as
\begin{align}
\Sigma \ &\equiv \ 
\frac{1}{\sqrt{2}} \ol{D}{}_+ D_- V
\nn \\
\ &= \ 
\sigma
+ \I \sqrt{2} \, \theta^+ \ol{\lambda}{}_+
- \I \sqrt{2} \, \ol{\theta}{}^- \lambda_-
- \sqrt{2} \, \theta^+ \ol{\theta}{}^- (D_V - \I F_{01})
+ \ldots
\, , \label{Sigma}
\end{align} 
where $F_{01} = \del_0 A_1 - \del_1 A_0$ is the field strength of the gauge field $A_m$.

We can easily construct an $\N=(2,2)$ supersymmetric Lagrangian in the following way:
\begin{align}
\Scr{L}_{\text{gauge}} 
\ &= \ 
\int \d^4 \theta \, \frac{1}{e^2} \Big( - |\Sigma|^2 + |\Phi|^2 \Big)
\, . \label{LGM}
\end{align}
Here we introduced the dimensionful gauge coupling constant $e$. 
We can also introduce two complex parameters 
$s \equiv \frac{1}{\sqrt{2}} (s^1 + \I s^2)$ and
$t \equiv \frac{1}{\sqrt{2}} (t^3 + \I t^4)$ 
in the following way:
\begin{align}
\Scr{L}_{\text{FI}} 
\ &= \ 
\Big\{ \sqrt{2} \int \d^2 \wt{\theta} \, t \, \Sigma
+ \text{(h.c.)} \Big\}
+ \Big\{ \sqrt{2} \int \d^2 \theta \, s \, \Phi
+ \text{(h.c.)} \Big\}
\, . \label{LFI}
\end{align}
We refer to $t$ as the complexified Fayet-Iliopoulos (FI) parameter. 
Indeed the FI term $t \Sigma$ gives rise to the $\N=(2,2)$ FI D-term and the topological term such as
$\int \d^2 \wt{\theta} \, t \Sigma + \text{(h.c.)} = - t^3 D_V - t^4 F_{01}$.
The other term $s \Phi$ in F-term is a natural $\N=(4,4)$ extension of the $\N=(2,2)$ FI term.

Imposing an invariance under the following exchange,
we can uplift (\ref{LGM}) to an $\N=(4,4)$ Lagrangian:
\begin{align}
(\sigma, \phi) \ \to \ (\sigma, \phi)
\, , \ls
(\lambda_{\pm}, \wt{\lambda}_{\pm}) \ \to \ 
(\wt{\lambda}_{\pm}, - \lambda_{\pm})
\, . \label{SU2R-gauge}
\end{align}
In section \ref{N44-symmetry} we will discuss an $\N=(4,4)$ extension with accuracy.

\subsection{Matter multiplets}

In addition to gauge multiplets, 
we can introduce matter multiplets (or called hypermultiplets).
In this article we introduce charged matter multiplets and neutral matter multiplets.
First we discuss a charged hypermultiplet involving
two complex scalars $(q, \wt{q})$,
four Weyl fermions $(\psi_{\pm}, \wt{\psi}_{\pm})$ with chirality $\pm$, 
and two complex auxiliary scalars $(F, \wt{F})$. 
This is also given in terms of $\N=(2,2)$ chiral superfields $Q$ and $\wt{Q}$ whose expansions are 
\bsubeq \label{SF-CHM}
\begin{align}
Q \ &= \ 
q 
+ \I \sqrt{2} \, \theta^+ \psi_+
+ \I \sqrt{2} \, \theta^- \psi_-
+ 2 \I \, \theta^+ \theta^- F
+ \ldots
\, , \label{Q} \\
\wt{Q} \ &= \ 
\wt{q} 
+ \I \sqrt{2} \, \theta^+ \wt{\psi}_+
+ \I \sqrt{2} \, \theta^- \wt{\psi}_-
+ 2 \I \, \theta^+ \theta^- \wt{F}
+ \ldots
\, . \label{tQ}
\end{align}
\esubeq
Now we assume that $Q$ (and $\wt{Q}$) has charge $+1$ (and $-1$) under $U(1)$ gauge symmetry.
A supersymmetric Lagrangian is constructed as
\begin{align}
\Scr{L}_{\text{CHM}}
\ &= \ 
\int \d^4 \theta \, \Big\{
|Q|^2 \, \e^{+2 V}
+ |\wt{Q}|^2 \, \e^{-2 V}
\Big\}
+ \Big\{ \sqrt{2} \int \d^2 \theta \, 
\Big( - \wt{Q} \Phi Q \Big)
+ \text{(h.c.)}
\Big\}
\, . \label{LCHM}
\end{align}
This is also uplifted to an $\N=(4,4)$ system if this is invariant under the following exchange:
\begin{align}
(q, \ol{\wt{q}}) \ \to \ (\ol{\wt{q}} , - q)
\, , \ls
(\psi_{\pm} , \ol{\wt{\psi}}{}_{\pm}) \ \to \ 
(\psi_{\pm} , \ol{\wt{\psi}}{}_{\pm}) 
\, . \label{SU2R-CHM}
\end{align}
In section \ref{N44-symmetry} we will rigorously discuss an $\N=(4,4)$ extension in terms of $SU(2)^3$ R-symmetry. 

We introduce a neutral matter multiplet, also referred to as a neutral hypermultiplet, which involves four real scalars $(r^1, r^2, r^3, r^4)$,
four Weyl fermions $(\chi_{\pm}, \wt{\chi}_{\pm})$,
and two complex auxiliary scalars $(G, \wt{G})$.
This multiplet is also given as a pair of $\N=(2,2)$ superfield $(\Psi, \Theta)$, where $\Psi$ is a chiral superfield and $\Theta$ is a twisted chiral superfield.
We explicitly write down their expansions,
\bsubeq \label{SF-NHM}
\begin{align}
\Psi
\ &= \ 
\frac{1}{\sqrt{2}} (r^1 + \I r^2)
+ \I \sqrt{2} \, \theta^+ \chi_+
+ \I \sqrt{2} \, \theta^- \chi_-
+ 2 \I \, \theta^+ \theta^- G
+ \ldots
\, , \label{Psi} \\
\Theta
\ &= \ 
\frac{1}{\sqrt{2}} (r^3 + \I r^4)
+ \I \sqrt{2} \, \theta^+ \ol{\wt{\chi}}{}_+
- \I \sqrt{2} \, \ol{\theta}{}^- \wt{\chi}_-
+ 2 \I \, \theta^+ \theta^- \wt{G}
+ \ldots
\, , \label{Theta} 
\end{align}
\esubeq
Introducing the dimensionless coupling constant $g$, we construct a Lagrangian of the neutral hypermultiplet,
\begin{align}
\Scr{L}_{\text{NHM}}
\ &= \ 
\int \d^4 \theta \, \frac{1}{g^2} \Big( - |\Theta|^2 + |\Psi|^2 \Big)
\, . \label{LNHM}
\end{align}
It is interesting to consider a coupling between the gauge multiplet $(V, \Phi)$ and the neutral hypermultiplet $(\Psi, \Theta)$ in the following form,
\begin{align}
\Scr{L}_{\text{FI2}} 
\ &= \ 
- \Big\{ \sqrt{2} \int \d^2 \wt{\theta} \, \Theta \Sigma
+ \text{(h.c.)} \Big\}
- \Big\{ \sqrt{2} \int \d^2 \theta \, \Psi \Phi
+ \text{(h.c.)} \Big\}
\, . \label{LFI2}
\end{align}
In the twisted F-term, $\Theta$ is topologically coupled to $\Sigma$.
The F-term is a natural $\N=(4,4)$ extension of the twisted F-term.
This coupling tells us that the FI parameters $(s,t)$ in (\ref{LFI}) can also be interpreted as expectation values of $(\Psi, \Theta)$, as well as the moment map of the gauge multiplet.
Due to the structure of the FI terms, 
we assume that the scalar $r^4$ is periodic with $2 \pi$ periodicity $r^4 \simeq r^4 + 2 \pi$,
while the other scalars $(r^1, r^2, r^3)$ take values in ${\mathbb R}^3$.
In section \ref{sect:defect5} we will study a geometrical meaning of the FI parameters.

\subsection{$\N=(4,4)$ R-symmetry}
\label{N44-symmetry}

We consider $\N=(4,4)$ theory whose building blocks are provided by $\N=(2,2)$ superfields.
In order to do so,
we introduce $SU(2) \times SO(4) \simeq SU(2)_1 \times SU(2)_2 \times SU(2)_3$ R-symmetry. 
The component fields are labeled by the following representations of the R-symmetry \cite{Tong:2002rq, Harvey:2005ab, Harvey:2014nha}:
\begin{gather}
\begin{array}{r@{\hspace{-1mm}}rl}
\multirow{2}{*}{$\text{gauge multiplet $(V, \Phi)$} \ \ \left\{ 
\begin{array}{c}
\vphantom{(\sigma, \phi)}
\cr
\vphantom{(\lambda_{\pm}, \wt{\lambda}_{\pm})}
\end{array}
\right.$}
&
(\sigma, \phi) 
\ &: \ \ 
({\bf 1}, {\bf 2}, {\bf 2})
\\
&
(\lambda_{\pm}, \wt{\lambda}_{\pm}) 
\ &: \ \ 
({\bf 2}, {\bf 2}, {\bf 1})_- \oplus ({\bf 2}, {\bf 1}, {\bf 2})_+
\\
\\
\multirow{2}{*}{$\text{charged matter multiplet $(Q, \wt{Q})$} \ \ \left\{ 
\begin{array}{c}
\vphantom{(q , \wt{q})}
\cr
\vphantom{(\psi_{\pm,}, \wt{\psi}_{\pm,})}
\end{array}
\right.$}
&
(q , \wt{q}) 
\ &: \ \ 
({\bf 2}, {\bf 1}, {\bf 1})
\\
&
(\psi_{\pm}, \wt{\psi}_{\pm}) 
\ &: \ \ 
({\bf 1}, {\bf 1}, {\bf 2})_- \oplus ({\bf 1}, {\bf 2}, {\bf 1})_+
\\
\\
\multirow{3}{*}{$\text{neutral matter multiplet $(\Psi, \Theta)$} \ \ \left\{ 
\begin{array}{c}
\vphantom{(r^1, r^2, r^3)}
\cr
\vphantom{r^4}
\cr
\vphantom{(\chi_{\pm}, \wt{\chi}_{\pm})}
\end{array}
\right.$}
&
(r^1, r^2, r^3)
\ &: \ \ 
({\bf 3}, {\bf 1}, {\bf 1})
\\
&
r^4 \
\ &: \ \ 
({\bf 1}, {\bf 1}, {\bf 1})
\\
&
(\chi_{\pm}, \wt{\chi}_{\pm})
\ &: \ \  
({\bf 2}, {\bf 1}, {\bf 2})_- \oplus ({\bf 2}, {\bf 2}, {\bf 1})_+
\end{array}
\label{SU2^3-charges}
\end{gather}
Here the subscripts $\pm$ in the right-hand side represent chirality.
The exchanges (\ref{SU2R-gauge}) and (\ref{SU2R-CHM}) are subject to the representations (\ref{SU2^3-charges}).
Indeed the Lagrangians (\ref{LGM}), (\ref{LFI}), (\ref{LCHM}), (\ref{LNHM}) and (\ref{LFI2}) are invariant under transformations by the above R-symmetry.

\section{Duality transformation formulae}
\label{sect:T-dual-F}

In this section we study two duality transformation formulae in two-dimensional supersymmetric theories. 
One is the formula of $\N=(2,2)$ (twisted) chiral superfields in D-terms based on \cite{Rocek:1991ps, Hori:2000kt}. 
The other is the formula of $\N=(2,2)$ chiral superfields\footnote{In this article we focus only on neutral chiral superfields.
In the case of charged chiral superfields, see \cite{Kimura:2014aja}.} in D-terms and F-terms \cite{Kimura:2013fda, Kimura:2014aja}. 
In particular, the latter formula is significant to study dualities among $\N=(4,4)$ supersymmetric theories which definitely contain F-terms in the language of $\N=(2,2)$ superfields.
They will play a central role in T-duality of string theory from the worldsheet point of view.

\subsection{Duality transformation in D-term}
\label{sect:T-dual-D}

First we demonstrate the well established formula.
We consider the duality transformation of the twisted chiral superfield $\Theta$ in (\ref{LNHM}) and (\ref{LFI2}).
Although $\Theta$ is in the twisted F-term of (\ref{LFI2}), this can be converted to a D-term by virtue of the definition of the twisted chiral superfield $\Sigma = \frac{1}{\sqrt{2}} \ol{D}{}_+ D_- V$.
Thus the $\Theta$ parts of the Lagrangians (\ref{LNHM}) and (\ref{LFI2}) can be described only in the D-term:
\begin{align}
\Scr{L}_{\Theta}
\ &= \
- \int \d^4 \theta \, \frac{1}{g^2} |\Theta|^2
- \Big\{ \sqrt{2} \int \d^2 \wt{\theta} \, \Theta \Sigma 
+ \text{(h.c.)} \Big\}
\nn \\
\ &= \ 
\int \d^4 \theta \, \Big\{
- \frac{1}{2 g^2} (\Theta + \ol{\Theta})^2 
- 2 (\Theta + \ol{\Theta}) V
\Big\}
+ \sqrt{2} \, \ve^{mn} \del_m (r^4 A_n)
\, , \label{LTheta-2}
\end{align}
where $\ve^{mn}$ is the Levi-Civita invariant tensor with normalization $\ve^{01} = +1$.
We now introduce an auxiliary real superfield $B$ and an auxiliary chiral superfield $\Gamma$ in (\ref{LTheta-2}),
\begin{align}
\Scr{L}_{\Theta B \Gamma}
\ &\equiv \ 
\int \d^4 \theta \, \Big\{
- \frac{1}{2 g^2} B^2
- 2 B V
- (\Gamma + \ol{\Gamma}) B
\Big\}
+ \sqrt{2} \, \ve^{mn} \del_m (r^4 A_n)
\, . \label{LThetaBGamma}
\end{align}
The original Lagrangian (\ref{LTheta-2}) can be realized if the auxiliary chiral superfield $\Gamma$ is integrated out.
This is because the equation of motion for $\Gamma$ is $0 = \ol{D}{}_+ \ol{D}{}_- B = D_+ D_- B$. Its solution is given by a sum of a twisted chiral superfield $\Theta$ such as $B = \Theta + \ol{\Theta}$.
Plugging this solution into (\ref{LThetaBGamma}), the original Lagrangian (\ref{LTheta-2}) appears.
On the other hand, if we integrate out the auxiliary real superfield $B$, we obtain a new system.
The solution of the equation of motion for $B$ is 
\begin{align}
\frac{1}{g^2} B \ &= \ - (\Gamma + \ol{\Gamma}) - 2 V
\, . \label{BGamma}
\end{align}
Substituting this into (\ref{LThetaBGamma}), the dual Lagrangian is given as
\begin{align}
\Scr{L}_{\Gamma} \ &\equiv \ 
\int \d^4 \theta \, 
\frac{g^2}{2} \Big( \Gamma + \ol{\Gamma} + 2 V \Big)^2
+ \sqrt{2} \, \ve^{mn} \del_m (r^4 A_n)
\, . \label{LGamma}
\end{align}
Here $\Gamma$ now becomes dynamical.
We notice that the power of the coupling constant $g$ is inverted.
Through the equations of motion for two superfields $\Gamma$ and $B$, 
we find the duality relation between the original dynamical twisted chiral superfield $\Theta$ and the new dynamical chiral superfield $\Gamma$ as follows:
\begin{align}
\frac{1}{g^2} (\Theta + \ol{\Theta})
\ &= \ 
- (\Gamma + \ol{\Gamma}) - 2 V
\, . \label{Theta2Gamma}
\end{align}
Expanding the chiral superfield $\Gamma$ in such a way as
\begin{align}
\Gamma \ &= \ 
\frac{1}{\sqrt{2}} (\gamma^3 + \I \gamma^4)
+ \I \sqrt{2} \, \theta^+ \zeta_+ 
+ \I \sqrt{2} \, \theta^- \zeta_-
+ 2 \I \, \theta^+ \theta^- G_{\Gamma}
+ \ldots
\, , \label{Gamma}
\end{align}
we can read off the duality relations among their dynamical component fields,
\bsubeq \label{Theta2Gamma-comp}
\begin{align}
r^3 \ &= \ - g^2 \gamma^3
\, , \\
\ol{\wt{\chi}}{}_{\pm} \ &= \ \mp g^2 \zeta_{\pm}
\, , \\
\pm (\del_0 \pm \del_1) r^4
\ &= \ 
- g^2 \big( (\del_0 \pm \del_1) \gamma^4 - \sqrt{2} (A_0 \pm A_1) \big)
\, .
\end{align}
\esubeq
It turns out that the scalar field $\gamma^4$ can be interpreted as a St\"{u}ckelberg field, i.e., this is gauge variant
such as $\gamma^4 \to \gamma^4 + \sqrt{2} \lambda$ under the gauge transformation $A_m \to A_m + \del_m \lambda$, although the original scalar $r^4$ is gauge invariant.
This feature is important for our consideration in section \ref{sect:defect5}.
Indeed this duality transformation formula is quite powerful.
This has been exhaustively utilized in the analysis of mirror symmetry in $\N=(2,2)$ systems \cite{Hori:2000kt}.

\subsection{Duality transformation in D-term and F-term}
\label{sect:T-dual-DF}

Second we perform a new formula discussed in \cite{Kimura:2013fda, Kimura:2014aja}.
We study the duality transformation of the chiral superfield $\Psi$ in (\ref{LNHM}) and (\ref{LFI2}).
Now $\Psi$ is coupled to the adjoint chiral superfield $\Phi$ in the $\N=(4,4)$ gauge multiplet. 
Since, by definition of the chiral superfield, $\Phi$ is given by
\begin{align}
\Phi \ &= \ \ol{D}{}_+ \ol{D}{}_- C
\end{align}
in terms of an unconstrained complex superfield $C$,
we can convert the F-term of (\ref{LFI2}) to D-terms.
Then the $\Psi$ parts of the Lagrangian (\ref{LNHM}) and (\ref{LFI2}) are given as
\begin{align}
\Scr{L}_{\Psi}
\ &= \ 
\int \d^4 \theta \, \frac{1}{g^2} |\Psi|^2 
- \Big\{ \sqrt{2} \int \d^2 \theta \, \Psi \Phi
+ \text{(h.c.)}
\Big\}
\nn \\
\ &= \ 
\int \d^4 \theta \, \Big\{
\frac{1}{2 g^2} (\Psi + \ol{\Psi})^2
- \sqrt{2} \, (\Psi + \ol{\Psi}) (C + \ol{C})
- \sqrt{2} \, (\Psi - \ol{\Psi}) (C - \ol{C})
\Big\}
\, . \label{LPsi-2}
\end{align}
We should notice that the system (\ref{LPsi-2}) contains the term $(\Psi - \ol{\Psi})(C - \ol{C})$, though any similar terms do not appear in (\ref{LTheta-2}) because $V$ is real.
Now we would like to perform a duality transformation by introducing auxiliary superfields as in (\ref{LThetaBGamma}).
In order to carry it out completely\footnote{The author thank Yutaka Matsuo, Shuhei Sasa, Yuji Tachikawa and Satoshi Watamura for pointing an incompleteness of this duality formula before publishing \cite{Kimura:2013fda}.}, we have to introduce two auxiliary real superfields $(R, S)$, two auxiliary twisted chiral superfields $(\Xi, \wt{\Xi})$, and an auxiliary chiral superfield $X$ in the following way:
\begin{align}
\Scr{L}_{RSX\Xi}
\ &\equiv \ 
\int \d^4 \theta \, \Big\{
\frac{1}{2 g^2} R^2 
- \sqrt{2} \, R (C + \ol{C})
+ R (\Xi + \ol{\Xi}) 
+ R (X + \ol{X})
\Big\}
\nn \\
\ & \ \ \ \ 
+ \int \d^4 \theta \, \Big\{
- \sqrt{2} \, (\I S) (C - \ol{C})
+ \I S (\wt{\Xi} - \ol{\wt{\Xi}})
+ \I S (X - \ol{X})
\Big\}
\, . \label{LRSXiX}
\end{align}

Let us first go back to the original Lagrangian (\ref{LPsi-2}) from (\ref{LRSXiX}).
Integrating out $(\Xi, \wt{\Xi})$ in the first step, we find
\bsubeq \label{sol-RS}
\begin{alignat}{3}
\ol{D}{}_+ D_- R \ &= \ D_+ \ol{D}{}_- R
&\ \ \ &\to& \ \ \
R \ &= \ 
\Psi_1 + \ol{\Psi}{}_1
\, , \label{sol-Xi} \\
\ol{D}{}_+ D_- (\I S) \ &= \ D_+ \ol{D}{}_- (\I S)
&\ \ \ &\to& \ \ \ 
\I S \ &= \ \Psi_2 - \ol{\Psi}{}_2
\, . \label{sol-tXi}
\end{alignat}
\esubeq
Here $\Psi_1$ and $\Psi_2$ are arbitrary chiral superfields.
Under these equations we integrate out $X$,
\bsubeq
\begin{align}
0 \ &= \ 
\ol{D}{}_+ \ol{D}{}_- (R + \I S) 
\ = \ 
\ol{D}{}_+ \ol{D}{}_- (\ol{\Psi}{}_1 - \ol{\Psi}{}_2)
\, , \\
0 \ &= \ 
D_+ D_- (R - \I S) 
\ = \ 
D_+ D_- (\Psi_1 - \Psi_2)
\, .
\end{align}
\esubeq 
The only one consistent solution which satisfies the above equations is
\begin{align}
\Psi_1 \ &= \ \Psi_2
\, . \label{Psi1=Psi2}
\end{align}
Plugging this into (\ref{sol-RS}) and (\ref{LRSXiX}), we obtain the same form as (\ref{LPsi-2}), where we can regard that the solution $\Psi_1$ is nothing but $\Psi$ in (\ref{LPsi-2}).

We go back to (\ref{LRSXiX}) and consider another configuration different from the original Lagrangian (\ref{LPsi-2}).
We integrate out the auxiliary twisted chiral superfield $\wt{\Xi}$ and the auxiliary real superfield $R$.
Each solution is given by (\ref{sol-tXi}) and 
\begin{align}
0 \ &= \ 
\frac{1}{g^2} R 
- \sqrt{2} (C + \ol{C})
+ (\Xi + \ol{\Xi})
+ (X + \ol{X})
\nn \\
\ &= \ 
\frac{1}{g^2} R
- \sqrt{2} (C' + \ol{C}{}')
+ (\Xi + \ol{\Xi})
\, . \label{sol-R}
\end{align}
Here we rewrote $\sqrt{2} C' = \sqrt{2} C - X$ without loss of generality, since both $C$ and $C'$ provide the same adjoint chiral superfield $\Phi$.
Substituting (\ref{sol-tXi}) and (\ref{sol-R}) into (\ref{LRSXiX}), 
we obtain the dual Lagrangian of (\ref{LPsi-2}),
\begin{align}
\Scr{L}_{\Xi} \ &\equiv \ 
\int \d^4 \theta \, \Big\{
- \frac{g^2}{2} \Big( \Xi + \ol{\Xi} - \sqrt{2} \, (C + \ol{C}) \Big)^2
- \sqrt{2} \, (\Psi - \ol{\Psi}) (C - \ol{C})
\Big\}
\, . \label{LXi}
\end{align}
Here we removed the prime attached with the superfield $C'$.
Now $\Xi$ becomes a new dynamical chiral superfield.
The duality relation between $\Psi$ and $\Xi$ can be discussed via the auxiliary fields $R$ and $\I S$ as
\bsubeq
\begin{align}
\Psi + \ol{\Psi} 
\ &= \ 
- g^2 (\Xi + \ol{\Xi})
+ \sqrt{2} \, g^2 (C + \ol{C})
\, , \label{Psi2Xi} \\
\Psi_1 \ &= \ \Psi_2 \ = \ \Psi
\, .
\end{align}
\esubeq
Expanding $\Xi$ and $C$ in such a way as
\bsubeq
\begin{align}
\Xi \ &= \ 
\frac{1}{\sqrt{2}} (y^1 + \I y^2)
+ \I \sqrt{2} \, \theta^+ \ol{\xi}{}_+
- \I \sqrt{2} \, \ol{\theta}{}^- \xi_-
+ 2 \I \, \theta^+ \ol{\theta}{}^- G_{\Xi}
+ \ldots 
\, , \label{Xi} \\
C \ &= \ 
\phi_{c} 
+ \I \sqrt{2} \, \theta^+ \psi_{c+} 
+ \I \sqrt{2} \, \theta^- \psi_{c-} 
+ \I \sqrt{2} \, \ol{\theta}{}^+ \chi_{c+} 
+ \I \sqrt{2} \, \ol{\theta}{}^- \chi_{c-}
\nn \\
\ & \ \ \ \ 
+ \I \, \theta^+ \theta^- F_{c} 
+ \I \, \ol{\theta}{}^+ \ol{\theta}{}^- M_{c}
+ \theta^+ \ol{\theta}{}^- G_{c} 
+ \theta^- \ol{\theta}{}^+ N_{c}
+ \theta^- \ol{\theta}{}^- A_{c=}
+ \theta^+ \ol{\theta}{}^+ B_{c\+}
\nn \\
\ & \ \ \ \ 
- \sqrt{2} \, \theta^+ \theta^- \ol{\theta}{}^+ \zeta_{c+}
- \sqrt{2} \, \theta^+ \theta^- \ol{\theta}{}^- \zeta_{c-}
- \sqrt{2} \, \theta^+ \ol{\theta}{}^+ \ol{\theta}{}^- \lambda_{c+}
- \sqrt{2} \, \theta^- \ol{\theta}{}^+ \ol{\theta}{}^- \lambda_{c-}
\nn \\
\ & \ \ \ \ 
- 2 \theta^+ \theta^- \ol{\theta}{}^+ \ol{\theta}{}^- D_{c}
\label{C}
\end{align}
\esubeq
with the relation among the component fields of $\Phi = \ol{D}{}_+ \ol{D}{}_- C$, 
\bsubeq \label{Phi-C}
\begin{align}
\phi
\ &= \ 
- \I \, M_{c} 
\, , \\
D_{\Phi}
\ &= \ 
- \I D_{c}
+ \frac{1}{2} (\del_0 - \del_1) B_{c\+}
+ \frac{1}{2} (\del_0 + \del_1) A_{c=}
+ \frac{\I}{2} (\del_0^2 - \del_1^2) \phi_{c}
\, , \\
\wt{\lambda}_{\pm}
\ &= \ 
- \I \Big\{ \lambda_{c\pm}
\pm (\del_0 \pm \del_1) \ol{\chi}{}_{c\mp}
\Big\}
\, , \\
&\ls
\{ \, 
F_{c} \, , \ \ 
G_{c} \, , \ \ 
N_{c} \, , \ \ 
\psi_{c\pm} \, , \ \ 
\zeta_{c\pm} \, \} \, : \ \ \ 
\text{(no relations)}
\, , 
\end{align}
\esubeq
we can also find the duality relations among their dynamical component fields,
\bsubeq \label{Psi2Xi-comp}
\begin{align}
r^1 \ &= \ 
- g^2 y^1 
+ g^2 (\phi_c + \ol{\phi}{}_c)
\, , \\
\chi_{\pm} \ &= \ 
\mp g^2 \ol{\xi}{}_{\pm}
+ \sqrt{2} \, g^2 (\psi_{c\pm} + \ol{\chi}{}_{c\pm})
\, , \\
(\del_0 + \del_1) r^2 
\ &= \ 
- g^2 (\del_0 + \del_1) y^2 
+ g^2 (B_{c\+} + \ol{B}{}_{c\+})
\, , \\
(\del_0 - \del_1) r^2 
\ &= \ 
+ g^2 (\del_0 - \del_1) y^2 
+ g^2 (A_{c=} + \ol{A}{}_{c=})
\, .
\end{align}
\esubeq

It seems strange that $\Psi - \ol{\Psi}$ exists in the dualized Lagrangian (\ref{LXi}) because this contains not only the original field $r^2$ but also its derivative $\del_m r^2$.
This would prevent the dynamical feature of the new field $y^2$.
In string worldsheet theory, the scalar field $y^2$ represents the physical coordinate while $r^2$ becomes the winding coordinate. 
We will carefully study their behaviors explicitly in section \ref{sect:defect5}.

Before ending this section, we have comments on the role of the auxiliary superfields in (\ref{LRSXiX}): 
\begin{itemize}
\item If $X$ is not introduced, one cannot find the coincidence (\ref{Psi1=Psi2}) which is essential to go back to the original Lagrangian (\ref{LPsi-2}).

\item If one integrates out the pair $(S, \Xi)$ instead of the pair $(R, \wt{\Xi})$, one obtains another dualized Lagrangian in which the real part of $\Psi$ is T-dualized.
This is intrinsically the same as (\ref{LXi}). 

\item If the pair $(R, S)$ is integrated out,
all component fields in $\Psi$ are mapped to those in $\Xi$.
This map does not imply the correct T-duality transformation from the viewpoint of the target space geometry in the IR limit.
\end{itemize}

\section{GLSMs for defect five-branes}
\label{sect:defect5}

In this section we study $\N=(4,4)$ GLSMs for defect five-branes \cite{Kimura:2013fda}. 
This is based on \cite{Tong:2002rq,Harvey:2005ab, Okuyama:2005gx}.
GLSM is a very powerful model because it can be regarded as the UV completion of string worldsheet sigma model.
In particular, because of the power of supersymmetry, 
$\N=(4,4)$ GLSM is also obtained by quiver gauge theory \cite{Douglas:1996sw, Johnson:1996py} and brane configuration \cite{Giveon:1998sr}, 
where $\N=(4,4)$ gauge theory is interpreted as an effective theory on D1-branes or compactified D2-branes.

We start from $\N=(4,4)$ $U(1)^k$ abelian gauge theory with $k$ gauge multiplets $(V_a, \Phi_a)$ coupled to $k$ charged hypermultiplets $(Q_a, \wt{Q}_a)$ and a neutral hypermultiplet $(\Psi, \Theta)$, where $a = 1,\ldots,k$.
We construct its Lagrangian whose constituents are (\ref{LGM}), (\ref{LFI}), (\ref{LCHM}), (\ref{LNHM}) and (\ref{LFI2}),
\begin{align}
\Scr{L}_1 \ &= \ 
\sum_{a=1}^k \int \d^4 \theta \, \Big\{
\frac{1}{e_a^2} \Big( - |\Sigma_a|^2 + |\Phi_a|^2 \Big)
+ |Q_a|^2 \, \e^{+2 V_a} 
+ |\wt{Q}_a|^2 \, \e^{-2 V_a}
\Big\}
+ \int \d^4 \theta \, \frac{1}{g^2} \Big( - |\Theta|^2 + |\Psi|^2 \Big)
\nn \\
\ & \ \ \ \ 
+ \sum_{a=1}^k \Big\{
\sqrt{2} \int \d^2 \theta \, \Big( 
- \wt{Q}_a \Phi_a Q_a
+ (s_a - \Psi) \Phi_a
\Big)
+ \text{(h.c.)}
\Big\}
\nn \\
\ & \ \ \ \ 
+ \sum_{a=1}^k \Big\{
\sqrt{2} \int \d^2 \wt{\theta} \, (t_a - \Theta) \Sigma_a
+ \text{(h.c.)}
\Big\}
\, . \label{GLSM-NS5}
\end{align}
This form looks quite generic in $\N=(4,4)$ theory.
In this section we analyze its low energy effective theory.
Indeed, the effective theory becomes a NLSM whose target space genuinely describes the transverse directions of NS5-branes \cite{Tong:2002rq}.
Furthermore, performing duality transformation formulae discussed in the previous section, 
we will obtain the GLSM for KK5-branes in section \ref{sect:GLSM-KK5}, and for exotic $5^2_2$-brane in section \ref{sect:GLSM-522}. 

Our strategy is as follows.
First, we study supersymmetric vacua.
We focus on the Higgs branch where the scalar fields of the neutral hypermultiplet take values in the non-trivial algebraic equations.
Second, we take the IR limit, where the dimensionful gauge coupling constant $e_a$ goes to infinity. 
Then the gauge multiplets become non-dynamical and we integrate them out.
After the integration we find a supersymmetric NLSM given by the neutral hypermultiplet.

\subsection{NS5-branes}
\label{sect:GLSM-NS5}

Let us first investigate the low energy effective theory of the GLSM (\ref{GLSM-NS5}).
We expand it in terms of the component fields,
\begin{align}
\Scr{L}_1 
\ &= \ 
\sum_a \frac{1}{e_a^2} \Big\{
\half (F_{01,a})^2 
- |\del_m \sigma_a|^2
- |\del_m \phi_a|^2
\Big\}
- \sum_a \Big\{
|D_m q_a|^2
+ |D_m \wt{q}_a|^2
\Big\}
\nn \\
\ & \ \ \ \ 
- \frac{1}{2 g^2}
\Big\{ (\del_m \vec{r})^2 + (\del_m r^4)^2 \Big\}
+ \sqrt{2} \sum_a (r^4 - t^4_a) \, F_{01,a}
\nn \\
\ & \ \ \ \ 
- 2 \sum_a \big( |\sigma_a|^2 + |\phi_a|^2 \big) 
\big( |q_a|^2 + |\wt{q}_a|^2 \big)
- 2 g^2 \sum_{a,b} \big( \sigma_a \ol{\sigma}{}_b + \phi_a \ol{\phi}{}_b \big)
\nn \\
\ & \ \ \ \ 
- \sum_a \frac{e_a^2}{2} 
\Big\{ |q_a|^2 - |\wt{q}_a|^2 - \sqrt{2} (r^3 - t^3_a) \Big\}^2
- \sum_a e_a^2 \Big|
\sqrt{2} \, q_a \wt{q}_a + \big( (r^1 - s^1_a) + \I (r^2 - s^2_a) \big) 
\Big|^2
\nn \\
\ & \ \ \ \ 
+ \text{(fermionic terms)}
\, , \label{GLSM-NS5-b1}
\end{align}
where $\vec{r}$ is a triplet of three scalar fields $\vec{r} = (r^1, r^2, r^3)$.
We have already integrated out all auxiliary fields in the supermultiplets.
We also introduced gauge covariant derivatives whose explicit forms are
\begin{align}
D_m q_a \ &= \ 
\del_m q_a - \I A_m \, q_a
\, , \ls
D_m \wt{q}_a \ = \ 
\del_m \wt{q}_a + \I A_m \, \wt{q}_a
\, . 
\end{align}
For simplicity, we ignore any fermionic terms.
The Lagrangian (\ref{GLSM-NS5-b1}) tells us the supersymmetric vacua in the following form,
\bsubeq \label{SUSYvacua-NS5}
\begin{align}
0 \ &= \
\big( |\sigma_a|^2 + |\phi_a|^2 \big) \big( |q_a|^2 + |\wt{q}_a|^2 \big)
\, , \\
0 \ &= \ 
\sum_{a,b} (\sigma_a \ol{\sigma}{}_b + \phi_a \ol{\phi}{}_b)
\, , \\
0 \ &= \ 
|q_a|^2 - |\wt{q}_a|^2 - \sqrt{2} (r^3 - t^3_a) 
\, , \\
0 \ &= \ 
\sqrt{2} \, q_a \wt{q}_a + \Big\{ (r^1 - s^1_a) + \I (r^2 - s^2_a) \Big\}
\, ,
\end{align}
\esubeq
whose solution on the Higgs branch is evaluated as
\bsubeq \label{sol-Higgs-NS5}
\begin{align}
0 \ &= \ \sigma_a \ = \ \phi_a
\, , \\
q_a \ &= \
\frac{\I}{2^{1/4}} \, \e^{+ \I \alpha_a} 
\sqrt{R_a + (r^3 - t^3_a)}
\, , \\
q_a \ &= \
\frac{\I}{2^{1/4}} \, \e^{- \I \alpha_a} 
\frac{(r^1 - s^1_a) + \I (r^2 - s^2_a)}{\sqrt{R_a + (r^3 - t^3_a)}}
\, , \\
R_a \ &= \
\half (|q_a|^2 + |\wt{q}_a|^2)
\ = \ 
\sqrt{(r^1 - s^1_a)^2 + (r^2 - s^2_a)^2 + (r^3 - t^3_a)^2}
\, ,
\end{align}
\esubeq
where $\alpha_a$ is a phase factor of the complex scalar field $q_a$.
Substituting (\ref{sol-Higgs-NS5}) into (\ref{GLSM-NS5-b1}), we obtain
\begin{align}
\Scr{L}_1
\ &= \ 
\sum_a \frac{1}{2 e_a^2} (F_{01,a})^2
- \half H (\del_m \vec{r})^2
- \frac{1}{2 g^2} (\del_m r^4)^2 
\nn \\
\ & \ \ \ \ 
- \sum_a \sqrt{2} R_a \Big( \del_m \alpha_a - A_{m,a} + \frac{1}{\sqrt{2}} \Omega_{i,a} \, \del_m r^i \Big)^2
+ \sqrt{2} \sum_a (r^4 - t^4_a) \, F_{01,a}
\nn \\
\ & \ \ \ \ 
+ \text{(fermionic terms)}
\, , \label{GLSM-NS5-b2}
\end{align}
with functions
\bsubeq \label{H-Omega}
\begin{align}
H \ &= \ 
\frac{1}{g^2} + \sum_a \frac{1}{\sqrt{2} R_a}
\, , \\
\Omega_{i,a} \, \del_m r^i
\ &= \ 
\frac{- (r^1 - s^1_a) \, \del_m r^2 + (r^2 - s^2_a) \, \del_m r^1}{\sqrt{2} R_a (R_a + (r^3 - t^3_a))}
\, . 
\end{align}
\esubeq
Consider the IR limit, where each gauge coupling constant $e_a$ goes to infinity $e_a \to \infty$ and the kinetic term of the gauge field disappears.
Then the gauge field becomes an auxiliary field.
Evaluating the field equation, the gauge field is solved as
\begin{align}
A_{m,a} \ &= \ 
\frac{1}{\sqrt{2}} \, \Omega_{i,a} \, \del_m r^i
- \frac{1}{2 R_a} \, \ve_{mn} \, \del^n (r^4 - t^4_a)
\, .
\end{align}
Here we have already used a gauge-fixing condition $\alpha_a = 0$.
Plugging this into the above Lagrangian under the IR limit, we obtain the bosonic part of the $\N=(4,4)$ NLSM,
\begin{align}
\Scr{L}_1^{\text{IR}} \ &= \ 
- \half H \Big\{ (\del_m \vec{r})^2 + (\del_m r^4)^2 \Big\}
+ \ve^{mn} \, \Omega_i \, \del_m r^i \, \del_n r^4
\, , \ls
\Omega_i \ \equiv \ 
\sum_a \Omega_{i,a}
\, . \label{NLSM-NS5}
\end{align}
We removed the derivative $\del^n t^4_a$ because $t^4_a$ is constant.
Compared this with the string worldsheet sigma model (\ref{string-NLSM-bosons}) with Table \ref{ST-WS}, we can read off the target space metric $G_{IJ}$ and B-field $B_{IJ}$ as follows:
\bsubeq \label{NS5-GB-NLSM}
\begin{alignat}{2}
G_{IJ} \ &= \ H \, \delta_{IJ}
\, , &\ls
I, J \ &= \ 6,7,8,9
\, , \\
B_{i9} \ &= \ \Omega_i
\, , &\ls
i \ &= \ 6,7,8
\, . 
\end{alignat}
We can also check that the functions $H$ and $\Omega_i$ satisfy the monopole equation 
\begin{align}
\nabla_i H \ &= \ (\nabla \times \vec{\Omega})_i
\, . 
\end{align}
\esubeq
Compared with the supergravity configuration (\ref{H-monopole}), 
we understand that the target space configuration (\ref{NS5-GB-NLSM}) represents the $k$-centered H-monopoles whose centers are labeled by the FI parameters $\vec{p}_a = (s^1_a, s^2_a, t^3_a)$ \cite{Okuyama:2005gx}.

So far we studied the two-dimensional theory which provides the string worldsheet sigma model of NS5-branes of codimension three.
We would also like to construct a sigma model for a defect NS5-brane of codimension two.
We deform the sigma model (\ref{NLSM-NS5}) to that of a defect NS5-brane via the smearing procedure \cite{deBoer:2010ud, Kikuchi:2012za, Kimura:2013fda}.
Now we compactify the $r^2$-direction on $S^1$ with radius ${\cal R}_8$.
The location of the H-monopoles in the $r^2$-direction becomes periodic
\begin{align}
s^2_a \ = \ 2 \pi {\cal R}_8 \, a
\, , \ls {a} \in {\mathbb Z}
\, . \label{r2-S1}
\end{align}
For simplicity, we set the H-monopoles in $r^1$- and $r^3$-directions to be at the origin $s^1_a = t^3_a = 0$.
In the small radius limit ${\cal R}_8 \to 0$,
we also introduce an infinite number of images of the H-monopoles as $k \to \infty$.
The discrete sum over $a$ is approximated by the continuous integral over $a$.
Thus we find \cite{Kimura:2013fda}
\bsubeq \label{Omega-kinfty}
\begin{gather}
H \ \xrightarrow{k\to\infty} \ 
H_{\sigma} \ \equiv \ h + \sigma \log \frac{\mu}{\varrho}
\, , \ls
\sigma \ \equiv \ \frac{1}{\sqrt{2} \, \pi {\cal R}_8}
\\
\Omega_1 \ \xrightarrow{k\to\infty} \ 0
\, , \ls
\Omega_2 \ \xrightarrow{k\to\infty} \ 
\Omega_{\sigma} 
\ \equiv \ 
\sigma \, \vartheta 
\, , \ls
\Omega_3 \ = \ 0
\, , \ls
\vartheta \ \equiv \ 
\arctan \Big( \frac{r^3}{r^1} \Big)
\, ,
\end{gather}
\esubeq
where $\varrho^2 = (r^1)^2 + (r^3)^2$.
Notice that the IR divergence has been regularized by the renormalization scale $\mu$, and $h$ is the ``bare'' quantity which diverges in the IR limit.
We stress that it is difficult to introduce the renormalization scale $\mu$ in the level of GLSM.
This is a reason why we gave up a direct construction of GLSMs for five-branes of codimension two. 
Instead, we started from the GLSM for $k$-centered five-branes of codimension three and performed the infinity limit $k \to \infty$ (\ref{Omega-kinfty}).
In this procedure the $r^2$-dependence of the functions $H$ and $\Omega_i$ disappears.
As a result, the $r^2$-direction is smeared and we obtain the background configuration of a defect NS5-brane codimension two,
whose target space configuration is given as
\begin{align}
G_{IJ} \ &= \ H_{\sigma} \, \delta_{IJ}
\, , \ls
B_{89} \ = \ \Omega_{\sigma}
\, . \label{dNS5-GB-NLSM}
\end{align}
This coincides with (\ref{dNS5}) in the supergravity framework.
Hence we conclude that the GLSM for a defect NS5-brane is constructed in the infinity limit $k \to \infty$ of the $\N=(4,4)$ supersymmetric $U(1)^k$ gauge theory with $k$ charged hypermultiplets with one neutral hypermultiplet.
We should keep in mind, however, that we have to take the infinity limit $k \to \infty$ after integrating out the gauge fields.
It implies that we can find the correct sigma model for defect five-branes if we take the infinity limit $k \to \infty$ {\it after} solving the field equations for the gauge fields in the IR limit $e_a \to \infty$,
while we cannot obtain it if we perform the infinity limit {\it before} the IR limit.
From now on we refer to the technique in (\ref{r2-S1}) and (\ref{Omega-kinfty}) as the ``smearing procedure''.

\subsection{KK5-branes}
\label{sect:GLSM-KK5}

Next we derive the NLSM for KK5-branes from $\N=(4,4)$ gauge theory \cite{Tong:2002rq}.
We go back to the Lagrangian (\ref{GLSM-NS5}).
Since the system of KK5-branes is T-dual of the system of H-monopoles, we should apply a duality transformation to the neutral hypermultiplet.
Dualizing the twisted chiral superfield $\Theta$ in (\ref{GLSM-NS5}) by using the formula discussed in section \ref{sect:T-dual-D}, we obtain
\begin{align}
\Scr{L}_2 \ &= \ 
\sum_{a=1}^k \int \d^4 \theta \, \Big\{
\frac{1}{e_a^2} \Big( - |\Sigma_a|^2 + |\Phi_a|^2 \Big)
+ |Q_a|^2 \, \e^{+2 V_a} 
+ |\wt{Q}_a|^2 \, \e^{-2 V_a}
\Big\}
\nn \\
\ & \ \ \ \ 
+ \int \d^4 \theta \, \Big\{ 
\frac{1}{g^2} |\Psi|^2 
+ \frac{g^2}{2} \Big( \Gamma + \ol{\Gamma} + 2 \sum_{a=1}^k V_a \Big)^2
\Big\}
\nn \\
\ & \ \ \ \ 
+ \sum_{a=1}^k \Big\{
\sqrt{2} \int \d^2 \theta \, \Big( 
- \wt{Q}_a \Phi_a Q_a
+ (s_a - \Psi) \Phi_a
\Big)
+ \text{(h.c.)}
\Big\}
\nn \\
\ & \ \ \ \ 
+ \sum_{a=1}^k \Big\{
\sqrt{2} \int \d^2 \wt{\theta} \, t_a \Sigma_a
+ \text{(h.c.)}
\Big\}
+ \sqrt{2} \, \ve^{mn} \sum_{a=1}^k \del_m (r^4 A_{n,a})
\, . \label{GLSM-KK5}
\end{align}
We investigate this low energy effective theory.
We focus on bosonic terms by expanding the superfields in the Lagrangian,
\begin{align}
\Scr{L}_2 \ &= \ 
\sum_a \frac{1}{e_a^2} \Big\{
\half (F_{01,a})^2 
- |\del_m \sigma_a|^2
- |\del_m \phi_a|^2
\Big\}
- \sum_a \Big\{
|D_m q_a|^2
+ |D_m \wt{q}_a|^2
\Big\}
\nn \\
\ & \ \ \ \ 
- \frac{1}{2 g^2} (\del_m \vec{r})^2
- \frac{g^2}{2} (D_m \gamma^4)^2
+ \sqrt{2} \, \ve^{mn} \sum_a \del_m \big( (r^4 - t^4_a) A_{n,a} \big)
\nn \\
\ & \ \ \ \ 
- 2 \sum_a \big( |\sigma_a|^2 + |\phi_a|^2 \big) 
\big( |q_a|^2 + |\wt{q}_a|^2 \big)
- 2 g^2 \sum_{a,b} \big( \sigma_a \ol{\sigma}{}_b + \phi_a \ol{\phi}{}_b \big)
\nn \\
\ & \ \ \ \ 
- \sum_a \frac{e_a^2}{2} \Big\{ |q_a|^2 - |\wt{q}_a|^2 - \sqrt{2} (r^3 - t^3_a) \Big\}^2
- \sum_a e_a^2 \Big| 
\sqrt{2} \, q_a \wt{q}_a + \big( (r^1 - s^1_a) + \I (r^2 - s^2_a) \big) 
\Big|^2
\nn \\
\ & \ \ \ \ 
+ \text{(fermionic terms)}
\, . \label{GLSM-KK5-b1}
\end{align}
Here we have already integrated out all auxiliary fields.
We introduced the covariant derivative of the St\"{u}ckelberg field $\gamma^4$ due to gauging of shift symmetry $\gamma^4 \to \gamma^4 + \sqrt{2} \lambda$ (see the duality relation (\ref{Theta2Gamma-comp})),
\begin{align}
D_m \gamma^4 \ &= \
\del_m \gamma^4 - \sqrt{2} \sum_a A_{m,a}
\, . \label{Dgamma4}
\end{align}
We extract the structure of supersymmetric vacua, which is the same as in (\ref{SUSYvacua-NS5}).
Then the Higgs branch of this system is also expressed by (\ref{sol-Higgs-NS5}).
Substituting (\ref{sol-Higgs-NS5}) into (\ref{GLSM-KK5-b1}), 
we find
\begin{align}
\Scr{L}_2 
\ &= \ 
\sum_a \frac{1}{2 e_a^2} (F_{01,a})^2
- \frac{1}{2} H (\del_m \vec{r})^2
- \frac{g^2}{2} (D_m \gamma^4)^2
\nn \\
\ & \ \ \ \ 
- \sum_a \sqrt{2} R_a \Big( \del_m \alpha_a - A_{m,a} + \frac{1}{\sqrt{2}} \Omega_{i,a} \del_m r^i \Big)^2
+ \sqrt{2} \, \ve^{mn} \sum_a \del_m \big( (r^4 - t^4_a) A_{n,a} \big)
\nn \\
\ & \ \ \ \ 
+ \text{(fermionic terms)}
\, , \label{GLSM-KK5-b2}
\end{align}
where $H$ and $\Omega_{i,a}$ are defined in (\ref{H-Omega}).

Consider the IR limit $e_a \to \infty$, where the gauge fields become non dynamical.
The solution of the field equation for each gauge field is 
\bsubeq \label{sol-A-KK5}
\begin{align}
A_{m,a} \ &= \
\frac{1}{2 R_a H} \Big( \del_m \wt{r}^4 - \Omega_i \, \del_m {r}^i \Big)
+ \del_m \alpha_a
+ \frac{1}{\sqrt{2}} \Omega_{i,a} \, \del_m {r}^i
\, , 
\end{align}
with introducing a gauge invariant variable $\wt{r}^4$,
\begin{align}
\wt{r}^4 \ &\equiv \ 
\gamma^4 - \sqrt{2} \sum_a \alpha_a
\, .
\end{align}
\esubeq
This is genuinely the dual field of the original scalar field $r^4$.
Substituting (\ref{sol-A-KK5}) into (\ref{GLSM-KK5-b2}) with gauge-fixing $\alpha_a = 0$ under the IR limit $e_a \to \infty$, we obtain 
\begin{align}
\Scr{L}_2^{\text{IR}} 
\ &= \ 
- \frac{1}{2} H (\del_m \vec{r})^2
- \half H^{-1} \Big( \del_m \wt{r}^4 - \Omega_i \, \del_m r^i \Big)^2
+ \sqrt{2} \, \ve^{mn} \sum_a \del_m \big( (r^4 - t^4_a) A_{n,a} \big) 
\nn \\
\ & \ \ \ \ 
+ \text{(fermionic terms)}
\, . \label{NLSM-KK5}
\end{align}
The total derivative term contains the gauge field $A_{n,a}$ subject to the solution (\ref{sol-A-KK5}).
Compared this with the solution (\ref{KK-monopole}) in the supergravity framework, 
it turns out that this is nothing but the NLSM for $k$-centered KK5-branes.
We note that the B-field is trivial up to a total derivative including the original scalar $r^4$, rather than the new dual scalar $\wt{r}^4$.
This form would be important when string winding charges is concerned \cite{Sen:1997zb, Gregory:1997te}, and evaluated by quantum corrections of GLSMs \cite{Tong:2002rq, Harvey:2005ab, Okuyama:2005gx}.
These days the winding charges can be discussed in double field theory \cite{Berman:2014jsa} and Alice string \cite{Okada:2014wma}.

We argue the configuration of a defect KK5-brane (or a KK-vortex) of codimension two.
In the same way as the defect NS5-brane, we perform the smearing procedure.
In the small radius limit ${\cal R}_8 \to 0$ as well as the infinity limit $k \to \infty$ under the relabeling of the FI parameter $s^2_a$ (\ref{r2-S1}), the functions $H$ and $\Omega_i$ in (\ref{NLSM-KK5}) are reduced to those in (\ref{Omega-kinfty}).
Plugging the result into (\ref{NLSM-KK5}), we find the sigma model for a defect KK5-brane which represents (\ref{dKK5}) from the supergravity viewpoint.
Again we claim that we should take the infinity limit $k \to \infty$ after the gauge multiplets are solved in the IR limit.
We cannot obtain the correct configuration of the defect KK5-brane if we take the infinity limit $k \to \infty$ before the integration of the gauge multiplets.

\subsection{Exotic $5^2_2$-brane}
\label{sect:GLSM-522}

In the previous subsection we performed the duality transformation and obtained the sigma model for (defect) KK5-branes.
If we perform dualization again, we will obtain a further dualized gauge theory.
Analyzing its low energy effective theory,
we will find the sigma model for an exotic $5^2_2$-brane of codimension two \cite{Kimura:2013fda}.
In this subsection we will write down many computations more explicitly than those in the previous subsections.
This is because we evaluate the role of unconstrained complex superfields $C_a$ (\ref{C}), which are not so frequently utilized in the literature.

We begin with the GLSM for KK5-branes (\ref{GLSM-KK5}) with the $\N=(4,4)$ neutral hypermultiplet $(\Psi, \Gamma)$.
We dualize the chiral superfield $\Psi$ to a twisted chiral superfield $\Xi$.
Following the discussion in section \ref{sect:T-dual-DF},
we obtain
\begin{align}
\Scr{L}_3 
\ &= \ 
\sum_{a=1}^k \int \d^4 \theta \, \Big\{
\frac{1}{e_a^2} \Big( - |\Sigma_a|^2 + |\Phi_a|^2 \Big)
+ |Q_a|^2 \, \e^{+ 2 V_a} 
+ |\wt{Q}_a|^2 \, \e^{- 2 V_a}
\Big\}
\nn \\
\ & \ \ \ \ 
+ \int \d^4 \theta \, \frac{g^2}{2} \Big\{
- \Big( \Xi + \ol{\Xi} - \sqrt{2} \sum_{a=1}^k (C_a + \ol{C}{}_a) \Big)^2
+ \Big( \Gamma + \ol{\Gamma} + 2 \sum_{a=1}^k V_a \Big)^2
\Big\}
\nn \\
\ & \ \ \ \ 
+ \sum_{a=1}^k \Big\{ 
\sqrt{2} \int \d^2 \theta \, \big( - \wt{Q}_a \Phi_a Q_a + s_a \, \Phi_a \big)
+ \text{(h.c.)}
\Big\}
+ \sum_{a=1}^k \Big\{ 
\sqrt{2} \int \d^2 \wt{\theta} \, t_a \, \Sigma_a
+ \text{(h.c.)}
\Big\}
\nn \\
\ & \ \ \ \ 
- \sqrt{2} \int \d^4 \theta \, (\Psi - \ol{\Psi}) \sum_{a=1}^k (C_a - \ol{C}{}_a)
+ \sqrt{2} \, \ve^{mn} \sum_{a=1}^k \del_m (r^4 A_{n,a})
\, . \label{GLSM-522}
\end{align}
Here the term $(\Psi - \ol{\Psi}) (C_a - \ol{C}{}_a)$ appears in this Lagrangian.
We emphasize that, although $\Psi$ is no longer dynamical after the dualization,
this term will play a crucial role in generating exotic functions. 
We express the Lagrangian (\ref{GLSM-522}) in terms of component fields,
\begin{align}
\Scr{L}_3 
\ &= \ 
\sum_a \frac{1}{e_a^2} \Big\{
\half (F_{01,a})^2
- |\del_m \sigma_a|^2
- |\del_m M_{c,a}|^2
\Big\}
- \sum_a \Big\{ 
|D_m q_a|^2
+ |D_m \wt{q}_a|^2
\Big\}
\nn \\
\ & \ \ \ \ 
- \frac{1}{2 g^2} \Big\{ (\del_m r^1)^2 + (\del_m r^3)^2 \Big\}
- \frac{g^2}{2} \Big\{ (\del_m y^2)^2 + (D_m \gamma^4)^2 \Big\}
+ \sqrt{2} \, \ve^{mn} \sum_a \del_m \big( (r^4 - t^4_a) A_{n,a} \big)
\nn \\
\ & \ \ \ \ 
- 2 g^2 \sum_{a,b} \sigma_a \ol{\sigma}{}_b 
- 2 \sum_a |\sigma_a|^2 \big( |q_a|^2 + |\wt{q}_a|^2 \big)
\nn \\
\ & \ \ \ \ 
+ \sum_a \Big\{ \frac{1}{2 e_a^2} (D_{V,a})^2
- D_{V,a} \big( |q_a|^2 - |\wt{q}_a|^2 - \sqrt{2} \, (r^3 - t^3_a) \big)
\Big\}
\nn \\
\ & \ \ \ \ 
+ \sum_a \Big\{ 
|F_a|^2 + |\wt{F}_a|^2
- \sqrt{2} M_{c,a} \big( q_a \wt{F}_a + \wt{q}_a F_a \big)
- \sqrt{2} \ol{M}{}_{c,a} \big( \ol{q}{}_a \ol{\wt{F}}{}_a + \ol{\wt{q}}{}_a \ol{F}{}_a \big)
\Big\}
+ g^2 |G_{\Gamma}|^2
\nn \\
\ & \ \ \ \ 
+ \frac{1}{\sqrt{2}} \sum_a \Big\{ (F_{c,a} - \ol{M}{}_{c,a}) \ol{G} + (\ol{F}{}_{c,a} - M_{c,a}) G \Big\}
- \frac{g^2}{2} \sum_{a,b} (F_{c,a} + \ol{M}{}_{c,a}) (\ol{F}{}_{c,b} + M_{c,b})
\nn \\
\ & \ \ \ \ 
+ {g^2} |{G}_{\Xi}|^2
+ \frac{\I g^2}{\sqrt{2}} \sum_a \Big\{ (G_{c,a} + \ol{N}{}_{c,a}) \ol{G}{}_{\Xi} - (\ol{G}{}_{c,a} + N_{c,a}) G_{\Xi} \Big\}
+ \frac{g^2}{2} \sum_{a,b} (G_{c,a} + \ol{N}{}_{c,a}) (\ol{G}{}_{c,b} + N_{c,b})
\nn \\
\ & \ \ \ \ 
+ \sum_a 
\frac{1}{e_a^2} 
\big| D_{c,a} - \sqrt{2} \, e_a^2 \, \ol{q}{}_a \ol{\wt{q}}{}_a \big|^2
- \sum_a 2 e_a^2 |q_a \wt{q}_a|^2
\nn \\
\ & \ \ \ \ 
- \sum_a D_{c,a} \Big\{ (r^1 - s^1_a) + \I (r^2 - s^2_a) \Big\}
- \sum_a \ol{D}{}_{c,a} \Big\{ (r^1 - s^1_a) - \I (r^2 - s^2_a) \Big\}
\nn \\
\ & \ \ \ \ 
- \sum_a \frac{\I}{2 e_a^2} 
\big( D_{c,a} - \sqrt{2} \, e_a^2 \, \ol{q}{}_a \ol{\wt{q}}{}_a \big)
\Big\{ (\del_0 - \del_1) \ol{B}{}_{c\+,a}
+ (\del_0 + \del_1) \ol{A}{}_{c=,a}
- \I (\del_0^2 - \del_1^2) \ol{\phi}{}_{c,a} \Big\}
\nn \\
\ & \ \ \ \ 
+ \sum_a \frac{\I}{2 e_a^2} 
\big( \ol{D}{}_{c,a} - \sqrt{2} \, e_a^2 \, q_a \wt{q}_a \big)
\Big\{ (\del_0 - \del_1) B_{c\+,a}
+ (\del_0 + \del_1) A_{c=,a}
+ \I (\del_0^2 - \del_1^2) \phi_{c,a} \Big\}
\nn \\
\ & \ \ \ \ 
+ \half \sum_a (\phi_{c,a} + \ol{\phi}{}_{c,a}) (\del_0^2 - \del_1^2) r^1 
+ \sum_a \frac{1}{4 e_a^2} \Big|
  (\del_0 - \del_1) B_{c\+,a}
+ (\del_0 + \del_1) A_{c=,a}
+ \I (\del_0^2 - \del_1^2) \phi_{c,a}
\Big|^2
\nn \\ 
\ & \ \ \ \  
- \frac{g^2}{2} \sum_a \Big\{
(B_{c\+,a} + \ol{B}{}_{c\+,a}) (\del_0 - \del_1) y^2
- (A_{c=,a} + \ol{A}{}_{c=,a}) (\del_0 + \del_1) y^2
\Big\}
\nn \\
\ & \ \ \ \ 
+ \frac{\I}{2} \sum_a 
(\phi_{c,a} - \ol{\phi}{}_{c,a}) (\del_0^2 - \del_1^2) r^2
- \frac{g^2}{2} \sum_{a,b} 
(A_{c=,a} + \ol{A}{}_{c=,a}) (B_{c\+,b} + \ol{B}{}_{c\+,b})
\nn \\
\ & \ \ \ \ 
+ \frac{\I}{2} \sum_a \Big\{
(B_{c\+,a} - \ol{B}{}_{c\+,a}) (\del_0 - \del_1) r^1
+ (A_{c=,a} - \ol{A}{}_{c=,a}) (\del_0 + \del_1) r^1 
\Big\}
\nn \\
\ & \ \ \ \ 
+ \text{(fermionic terms)}
\, . \label{GLSM-522-b1}
\end{align}
We now evaluate the field equations for the auxiliary fields 
$D_{V,a}$, $F_a$, $\wt{F}_a$, 
$G_{\Gamma}$, $G$, $F_{c,a}$, 
$G_{\Xi}$, $G_{c,a}$, $N_{c,a}$,
$D_{c,a}$, $A_{c=,a}$, $B_{c\+,a}$ and $\phi_{c,a}$ respectively,
\bsubeq \label{eom-aux-522}
\begin{align}
0 \ &= \ 
\frac{1}{e_a^2} D_{V,a}
- \Big\{ |q_a|^2 - |\wt{q}_a|^2 - \sqrt{2} \, (r^3 - t^3_a) \Big\}
\, , \\
0 \ &= \ 
\ol{F}{}_a 
- \sqrt{2} \, M_{c,a} \, \wt{q}_a 
\, , \\
0 \ &= \ 
\ol{\wt{F}}{}_a
- \sqrt{2} \, M_{c,a} \, q_a
\, , \\
0 \ &= \
\ol{G}{}_{\Gamma}
\, , \\
0 \ &= \
\sum_b (\ol{F}{}_{c,b} - M_{c,b})
\, , \\
0 \ &= \
\ol{G}
- \frac{g^2}{\sqrt{2}} \sum_b (\ol{F}{}_{c,b} + M_{c,b})
\, , \\
0 \ &= \
\ol{G}{}_{\Xi}
- \frac{\I}{\sqrt{2}} \sum_b (\ol{G}{}_{c,b} + N_{c,b})
\, , \\
0 \ &= \
\ol{G}{}_{\Xi}
- \frac{\I}{\sqrt{2}} \sum_b (\ol{G}{}_{c,b} + N_{c,b})
\, , \\
0 \ &= \
G_{\Xi}
+ \frac{\I}{\sqrt{2}} \sum_b (G_{c,b} + \ol{N}{}_{c,b})
\, , \\
0 \ &= \ 
\frac{1}{e_a^2} \big( \ol{D}{}_{c,a} - \sqrt{2} \, e_a^2 \, q_a \wt{q}_a \big)
- \frac{\I}{2 e_a^2} \Big[
  (\del_0 - \del_1) \ol{B}{}_{c\+,a}
+ \I (\del_0 + \del_1) \ol{A}{}_{c=,a}
- \I (\del_0^2 - \del_1^2) \ol{\phi}{}_{c,a}
\Big]
\nn \\
\ & \ \ \ \ 
- \Big\{ (r^1 - s^1_a) + \I (r^2 - s^2_a) \Big\}
\, , \label{EOM-Dca} \\
0 \ &= \ 
(\del_0 + \del_1) \Big\{
- \frac{\I}{e_a^2} \big( \ol{D}{}_{c,a} - \sqrt{2} \, e_a^2 q_a \wt{q}_a \big)
- \frac{1}{2 e_a^2} \Big[ (\del_0 - \del_1) \ol{B}{}_{c\+,a} + (\del_0 + \del_1) \ol{A}{}_{c=,a} - \I (\del_0^2 - \del_1^2) \ol{\phi}{}_{c,a} \Big]
\Big\}
\nn \\
\ & \ \ \ \ 
+ \I (\del_0 + \del_1) r^1 
+ g^2 (\del_0 + \del_1) y^2
- g^2 \sum_b (B_{c\+,b} + \ol{B}{}_{c\+,b})
\, , \label{EOM-Ac} \\
0 \ &= \ 
(\del_0 - \del_1) \Big\{
- \frac{\I}{e_a^2} \big( \ol{D}{}_{c,a} - \sqrt{2} \, e_a^2 q_a \wt{q}_a \big)
- \frac{1}{2 e_a^2} \Big[ (\del_0 - \del_1) \ol{B}{}_{c\+,a} + (\del_0 + \del_1) \ol{A}{}_{c=,a} - \I (\del_0^2 - \del_1^2) \ol{\phi}{}_{c,a} \Big]
\Big\}
\nn \\
\ & \ \ \ \ 
+ \I (\del_0 - \del_1) r^1
- g^2 (\del_0 - \del_1) y^2
- g^2 \sum_b (A_{c=,b} + \ol{A}{}_{c=,b})
\, , \label{EOM-Bc} \\
0 \ &= \ 
(\del_0^2 - \del_1^2) \Big\{
\frac{1}{e_a^2} \big( \ol{D}{}_{c,a} - \sqrt{2} \, e_a^2 q_a \wt{q}_a \big)
- \frac{\I}{2 e_a^2} \Big[ (\del_0 - \del_1) \ol{B}{}_{c\+,a} + (\del_0 + \del_1) \ol{A}{}_{c=,a} - \I (\del_0^2 - \del_1^2) \ol{\phi}{}_{c,a} \Big]
\Big\}
\nn \\
\ & \ \ \ \ 
- (\del_0^2 - \del_1^2) (r^1 + \I r^2)
\, . \label{EOM-phic}
\end{align}
\esubeq
The equations for $A_{c=,a}$ (\ref{EOM-Ac}), $B_{c\+,a}$ (\ref{EOM-Bc}), and $\phi_{c,a}$ (\ref{EOM-phic}) are trivial under the equation for $D_{c,a}$ (\ref{EOM-Dca}) and the duality relations (\ref{Psi2Xi-comp}).
Plugging (\ref{eom-aux-522}) into the Lagrangian (\ref{GLSM-522-b1}),
we obtain a very clear description,
\begin{align}
\Scr{L}_3 
\ &= \ 
\sum_a \frac{1}{e_a^2} \Big\{
\half (F_{01,a})^2 
- |\del_m \sigma_a|^2
- |\del_m M_{c,a}|^2
\Big\}
- \sum_a \Big\{ 
|D_m q_a|^2
+ |D_m \wt{q}_a|^2
\Big\}
\nn \\
\ & \ \ \ \ 
- \frac{1}{2 g^2} \Big\{ (\del_m r^1)^2 + (\del_m r^3)^2 \Big\}
- \frac{g^2}{2} \Big\{ (\del_m y^2)^2 + (D_m \gamma^4)^2 \Big\}
+ \sqrt{2} \, \ve^{mn} \sum_a \del_m \big( (r^4 - t^4_a) A_{n,a} \big)
\nn \\
\ & \ \ \ \ 
- 2 g^2 \sum_{a,b} \big( \sigma_a \ol{\sigma}{}_b + M_{c,a} \ol{M}{}_{c,b} \big)
- 2 \sum_a \big( |\sigma_a|^2 + |M_{c,a}|^2 \big) 
\big( |q_a|^2 + |\wt{q}_a|^2 \big)
\nn \\
\ & \ \ \ \ 
- \sum_a \frac{e_a^2}{2} \Big\{ |q_a|^2 - |\wt{q}_a|^2 - \sqrt{2} \, (r^3 - t^3_a) \Big\}^2
- \sum_a e_a^2 \Big| \sqrt{2} \, q_a \wt{q}_a + \big( (r^1 - s^1_a) + \I (r^2 - s^2_a) \big) \Big|^2
\nn \\
\ & \ \ \ \ 
+ \frac{g^2}{2} \sum_{a,b} 
(A_{c=,a} + \ol{A}{}_{c=,a}) (B_{c\+,b} + \ol{B}{}_{c\+,b})
\nn \\
\ & \ \ \ \ 
+ \text{(fermionic terms)}
\, . \label{GLSM-522-b2}
\end{align}
We emphasize that not only the new dynamical scalar $y^2$ but also the original ``non-dynamical'' scalar field $r^2$ are involved in the system.
The latter is interpreted as an ``auxiliary'' field in the present stage.
Furthermore, we note that the term $(A_{c=,a} + \ol{A}{}_{c=,a}) (B_{c\+,b} + \ol{B}{}_{c\+,b})$ is subject to the duality relation (\ref{Psi2Xi-comp}).
This will play an important role in the next analysis.

We investigate the structure of the Higgs branch.
In order that a vacuum is supersymmetric, we should impose a set of constraints,
\bsubeq \label{SUSYvacua-522}
\begin{align}
0 \ &= \ 
\sigma_a \ = \ M_{c,a}
\, , \\
0 \ &= \ 
|q_a|^2 - |\wt{q}_a|^2 - \sqrt{2} \, (r^3 - t^3_a)
\, , \\
0 \ &= \ 
\sqrt{2} \, q_a \wt{q}_a + \big( (r^1 - s^1_a) + \I (r^2 - s^2_a) \big)
\, , \\
0 \ &= \ 
\frac{g^2}{2} \sum_{a,b} (A_{c=,a} + \ol{A}{}_{c=,a}) (B_{c\+,b} + \ol{B}{}_{c\+,b})
\nn \\
\ &= \ 
- \frac{1}{2 g^2} (\del_m r^2)^2
+ \frac{g^2}{2} (\del_m y^2)^2 
+ \ve^{mn} (\del_m r^2) (\del_n y^2)
\, . \label{AB-const}
\end{align}
\esubeq
The second and third constraints provide the same solution of two charged scalar fields $q_a$ and $\wt{q}_a$ as in (\ref{sol-Higgs-NS5}).
On the other hand, the last equation (\ref{AB-const}) indicates that the new dual field $y^2$ and the ``auxiliary'' field $r^2$ are strongly related to each other under the duality relation (\ref{Psi2Xi-comp}).
Substituting the solutions in (\ref{GLSM-522-b2}),
we obtain 
\begin{align}
\Scr{L}_3
\ &= \ 
\sum_a \frac{1}{2 e_a^2} (F_{01,a})^2
- \frac{1}{2} H \Big\{ (\del_m r^1)^2 + (\del_m r^2)^2 + (\del_m r^3)^2 \Big\}
- \frac{g^2}{2} (D_m \gamma^4)^2
\nn \\
\ & \ \ \ \ 
- \sum_a \sqrt{2} R_a \Big( \del_m \alpha_a - A_{m,a} + \frac{1}{\sqrt{2}} \Omega_{i,a} \, \del_m r^i \Big)^2
\nn \\
\ & \ \ \ \ 
+ \ve^{mn} \, (\del_m r^2) (\del_n y^2)
+ \sqrt{2} \, \ve^{mn} \sum_a \del_m \big( (r^4 - t^4_a) A_{n,a} \big)
\nn \\
\ & \ \ \ \ 
+ \text{(fermionic terms)}
\, . \label{GLSM-522-b3}
\end{align}
The functions $H$ and $\Omega_{i,a}$ are defined in (\ref{H-Omega}).
It is quite interesting that the kinetic term of 
the ``auxiliary'' field $r^2$ {\it revives},
whilst that of the dynamical field $y^2$ {\it disappears}.
However, we should understand that $y^2$ is genuinely dynamical and $r^2$ is non-dynamical via the constraint (\ref{AB-const}).
Indeed, this phenomenon originates from $(\Psi - \ol{\Psi}) (C_a - \ol{C}{}_a)$ in (\ref{GLSM-522}), 
and the ``second derivative'' term $H (\del_m r^2)^2$ in (\ref{GLSM-522-b3}) will contribute to the final form of NLSM.

We study the low energy theory in the IR limit $e_a \to \infty$, where the gauge fields $A_{m,a}$ are non dynamical.
Each solution of the field equations for $A_{m,a}$ is the same as (\ref{sol-A-KK5}).
Plugging it in (\ref{GLSM-522-b3}) under a gauge-fixing condition $\alpha_a = 0$, 
we find
\begin{align}
\Scr{L}_3^{\text{IR}}  
\ &= \ 
- \frac{1}{2} H \Big\{ (\del_m r^1)^2 + (\del_m r^2)^2 + (\del_m r^3)^2 \Big\}
- \frac{1}{2 H} (\del_m \wt{r}^4)^2 
- \frac{(\Omega_2)^2}{2 H} (\del_m r^2)^2
+ \frac{\Omega_2}{H} (\del_m r^2) (\del^m \wt{r}^4) 
\nn \\
\ & \ \ \ \ 
- \frac{(\Omega_1)^2}{2 H} (\del_m r^1)^2
- \frac{\Omega_1 \Omega_2}{H} 
(\del_m r^1) (\del^m r^2)
+ \frac{\Omega_1}{H} (\del_m r^1) (\del^m \wt{r}^4) 
\nn \\
\ & \ \ \ \ 
+ \ve^{mn} (\del_m r^2) (\del_n y^2) 
+ \sqrt{2} \, \ve^{mn} \sum_a \del_m \big( (r^4 - t^4_a) A_{n,a} \big)
\nn \\
\ & \ \ \ \ 
+ \text{(fermionic terms)}
\, . \label{GLSM-522-b4}
\end{align}
The total derivative term contains the gauge field $A_{n,a}$ subject to the solution (\ref{sol-A-KK5}).
This is not the final form because we have to integrate out the ``auxiliary'' field $r^2$.
Before doing that, we perform the smearing procedure to make a shift symmetry $r^2 \to r^2 + \text{(constant)}$.
Applying the infinity limit $k \to \infty$ with the small radius limit ${\cal R}_8 \to 0$ to the system (\ref{GLSM-522-b4}), we obtain the reduction (\ref{Omega-kinfty}) and the Lagrangian is also reduced to
\bsubeq \label{GLSM-522-b5}
\begin{align}
\Scr{L}_3^{\text{IR}} 
\ &= \ 
- \frac{1}{2} H_{\sigma} \Big\{ (\del_m \varrho)^2 + \varrho^2 (\del_m \vartheta)^2 \Big\}
- \frac{1}{2 H_{\sigma}} (\del_m \wt{r}^4)^2 
\nn \\
\ & \ \ \ \ 
- \frac{K_{\sigma}}{2 H_{\sigma}} (\del_m r^2)^2
+ \frac{\Omega_{\sigma}}{H_{\sigma}} (\del_m r^2) (\del^m \wt{r}^4) 
+ \ve^{mn} (\del_m r^2) (\del_n y^2) 
+ \sqrt{2} \, \ve^{mn} \sum_a \del_m \big((r^4 - t^4_a) A_{n,a} \big)
\nn \\
\ & \ \ \ \ 
+ \text{(fermionic terms)}
\, , \\
r^1 \ &\equiv \ \varrho \, \cos \vartheta
\, , \ls
r^3 \ \equiv \ \varrho \, \sin \vartheta
\, , \ls
K_{\sigma} \ \equiv \ (H_{\sigma})^2 + (\Omega_{\sigma})^2
\, . 
\end{align}
\esubeq
Thus we are now able to integrate out $r^2$ without any difficulty since there are no non-derivative term of it.
Integration over $r^2$ also generates the correct kinetic term of $y^2$ due to the topological term $\ve^{mn} (\del_m r^2) (\del_n y^2)$ in (\ref{GLSM-522-b5}).
Hence we finally obtain 
\begin{align}
\Scr{L}_3^{\text{IR}}  
\ &= \ 
- \frac{1}{2} H_{\sigma} \Big\{ (\del_m \varrho)^2 + \varrho^2 (\del_m \vartheta)^2 \Big\}
- \frac{H_{\sigma}}{2 K_{\sigma}} \Big\{ (\del_m y^2)^2 + (\del_m \wt{r}^4)^2 \Big\}
- \frac{\Omega_{\sigma}}{K_{\sigma}} \, \ve^{mn} (\del_m y^2) (\del_n \wt{r}^4)
\nn \\
\ & \ \ \ \ 
+ \sqrt{2} \, \ve^{mn} \sum_a \del_m \big((r^4 - t^4_a) A_{n,a} \big)
+ \text{(fermionic terms)}
\, . \label{NLSM-522}
\end{align}
Compared with the background configuration (\ref{522}) in the supergravity framework, we conclude that (\ref{NLSM-522}) is nothing but the string sigma model for an exotic $5^2_2$-brane.
We notice that the constraint (\ref{AB-const}) is indeed crucial to generate the function $H_{\sigma} / K_{\sigma}$ in front of the second derivative of $y^2$.
In other words, we understand that the ``auxiliary'' field $r^2$ in this dual system plays an significant role in constructing the string sigma model for the exotic five-brane.
We should also comment that the physical meaning of $r^2$ from the target space viewpoint is nothing but the winding coordinate in the $5^2_2$-brane system.
Integrating it out, the target space configuration of (\ref{NLSM-522}) is governed by the multi-valued function $K_{\sigma}$ under the shift $\vartheta \to \vartheta + 2 \pi$.
If we do not integrate it out in the Lagrangian (\ref{GLSM-522-b5}),
the target space coordinate is {\it doubled} from $y^2$ to $(y^2, r^2)$, and the configuration is single-valued with respect to all variables \cite{Okada:2014wma}.
Then we can say that the NLSM (\ref{GLSM-522-b5}) is a doubled sigma model \cite{Hull:2004in, Hull:2006va, Dall'Agata:2008qz, Berman:2014jsa} for the exotic $5^2_2$-brane.
We also comment that the doubled structure can be traced by $\beta$-supergravity \cite{Andriot:2013xca, Andriot:2014uda, Sakatani:2014hba}.

\section{Summary and discussions}
\label{sect:summary}


In this article we reviewed two-dimensional $\N=(4,4)$ GLSMs which provide string worldsheet sigma models for (defect) five-branes.
First we considered the generic rule of duality transformations in two-dimensional supersymmetric theories in the superfield formalism.
In particular, we discussed the duality transformation in the presence of F-terms as well as D-terms.
Next we concretely analyzed $\N=(4,4)$ GLSMs for five-branes of codimension three, and for defect five-branes of codimension two.
In order to realize exotic feature of defect five-branes, we performed the smearing procedure, in which we took the infinity limit of the number of gauge multiplets $k$ after the IR limit.
The duality transformation in the presence of F-terms also played the central role in constructing the GLSM for the exotic $5^2_2$-brane.
Analyzing the gauge theory whose IR limit is the string worldsheet sigma model for the $5^2_2$-brane, we understood that the exotic feature originates from the string winding coordinate field $r^2$.
This is because that the configuration becomes multi-valued when we integrate out $r^2$, while this is still single-valued if we do not integrate it out.
We also found that the sigma model containing both $y^2$ and $r^2$ is a doubled sigma model.
This is a typical example of T-fold and nongeometric string backgrounds \cite{Hull:2004in, Hull:2006va, Berman:2014jsa}.
Thus we concluded that the $\N=(4,4)$ GLSM for the exotic $5^2_2$-brane (\ref{GLSM-522}) is a very powerful model beyond supergravity analyses, when we explore exotic features in string theory.


We would like to discuss various analyses after we established the work \cite{Kimura:2013fda}.
Just after constructing the GLSM \cite{Kimura:2013fda},
we investigated quantum corrections to the string sigma model by virtue of quantum vortex corrections in gauge theory.
This was motivated by the works \cite{Tong:2002rq, Harvey:2005ab, Okuyama:2005gx} in which the vortex corrections can be interpreted as the worldsheet instanton corrections to string sigma model.
In particular, the vortex corrections in the $\N=(4,4)$ GLSM break the isometry of the target space.
We studied the vortex corrections in gauge theory and obtained the winding corrections of the configuration of the exotic $5^2_2$-brane \cite{Kimura:2013zva, Kimura:2013khz}. 
This result is given by the modified Bessel function of the second kind, which also appears in the study of D-instanton corrections to the moduli space of hypermultiplets in Calabi-Yau compactifications \cite{Ooguri:1996me}.

We also studied the worldvolume actions of exotic $5^2_2$-branes \cite{Chatzistavrakidis:2013jqa, Kimura:2014upa} based on the work by Bergshoeff and his company \cite{Bergshoeff:1997gy}.
We constructed the Dirac-Born-Infeld actions of exotic $5^2_2$-branes in type IIA and IIB string theories. 
They are fully covariant with respect to two Killing vectors associated with two isometries. 
The effective theories are governed by the six-dimensional $\N = (2, 0)$ tensor multiplet in type IIA string, and $\N = (1, 1)$ vector multiplet in type IIB string.

It is also important to understand bound states of various defect five-branes, i.e., defect $(p,q)$ five-branes \cite{Kimura:2014wga, Kimura:2014bea}.
They would tell us the global structure of exotic five-branes \cite{Hellerman:2002ax, Kikuchi:2012za}. 
In the current stage we recognize it only from that of seven-branes via string dualities \cite{Greene:1989ya, Bergshoeff:2006jj}.
Our next research is to derive it only in the playground of five-branes without the aid of seven-branes.

\section*{Acknowledgements}

The author thank
Yuho Sakatani,
Shin Sasaki, 
Masaki Shigemori
and 
Masaya Yata 
for exciting collaborations and fruitful discussions.
He is also grateful to
Machiko Hatsuda,
Yosuke Imamura,
Yusuke Kimura,
Yutaka Matsuo,
Shun'ya Mizoguchi,
Takahiro Nishinaka,
Kazutoshi Ohta,
Shuhei Sasa,
Yuji Tachikawa,
Masato Taki,
Seiji Terashima,
Satoshi Watamura,
Satoshi Yamaguchi,
Daisuke Yokoyama
and
Yutaka Yoshida
for valuable discussions and verbal encouragement. 
He expresses his gratitude to Yukawa Institute for Theoretical Physics at
Kyoto University. Discussions during the YITP molecule-type workshop on
``{\sl Exotic Structures of Spacetime}'' ({\tt YITP-T-13-07}) are helpful to his current research.
This work is supported in part by Iwanami-Fujukai Foundation.

\begin{appendix}

\section*{Appendix}

\section{Conventions}
\label{app:conventions}

\subsection{2D Lorentz signature}

First of all, we introduce two vector indices in the following way:
\begin{align*}
m,n,p, \ldots = 0,1 &: \ \ \text{indices of 2D curved spacetime}
\, , \\
a,b,c, \ldots = \hat{0}, \hat{1} &: \ \ \text{indices of 2D tangent spacetime}
\, .
\end{align*}
In this article we adopt the Lorentz signature $g_{mn} = {\rm diag}(-,+)$ in two-dimensional spacetime.

\subsection{2D superspace}

We introduce two-dimensional superspace expanded by the conventional coordinates $x^m$ and the anti-commuting Grassmann coordinates $\theta^{\alpha}$ and $\ol{\theta}{}^{\dot{\alpha}} = (\theta^{\alpha})^{\dagger}$. They are Weyl spinors.
The feature is summarized as
\bsubeq
\begin{gather}
(\theta^1, \theta^2) \ \equiv \ (\theta^-, \theta^+)
\, , \ls
(\theta^{\alpha})^{\dagger} \ = \ \ol{\theta}{}^{\dot{\alpha}}
\, , \ls
(\theta^{\pm})^{\dagger} \ = \ \ol{\theta}{}^{\pm}
\, , \\
\theta_{\alpha} \ = \ \ve_{\alpha \beta} \theta^{\beta}
\, , \ls
\theta^{\alpha} \ = \ \ve^{\alpha \beta} \theta_{\beta}
\, , \\
\ve^{-+} \ = \ \ve_{+-} \ = \ + 1
\, , \ls
\theta^- \ = \ + \theta_+ 
\, , \ls
\theta^+ \ = \ - \theta_-
\, , \\
\begin{align}
2 \theta^+ \theta^- 
\ &= \ 
+ \theta^+ \theta^- - \theta^- \theta^+
\ = \ 
\theta^{\alpha} (\ve_{\alpha \beta} \theta^{\beta})
\ = \ 
\theta^{\alpha} \theta_{\alpha}
\ \equiv \ 
\theta \theta 
\, , \\
- 2 \ol{\theta}{}^+ \ol{\theta}{}^- 
\ &= \ 
+ \ol{\theta}{}^- \ol{\theta}{}^+ - \ol{\theta}{}^+ \ol{\theta}{}^-
\ = \ 
- \ol{\theta}{}_+ \ol{\theta}{}_- + \ol{\theta}{}_- \ol{\theta}{}_+
\ = \ 
\ol{\theta}{}_{\dot{\alpha}} ( \ve^{\dot{\alpha} \dot{\beta}} \ol{\theta}{}_{\dot{\beta}})
\ = \ 
\ol{\theta}{}_{\dot{\alpha}} \ol{\theta}{}^{\dot{\alpha}}
\ \equiv \ 
\ol{\theta} \ol{\theta} 
\, .
\end{align}
\end{gather}
\esubeq
We also introduce supercovariant derivatives $D_{\alpha}$, $\ol{D}{}_{\dot{\alpha}}$ and the derivative representation of supercharges $Q_{\alpha}$, $\ol{Q}{}_{\dot{\alpha}}$ as follows:
\bsubeq \label{DQ}
\begin{align}
D_{\pm} \ &= \ 
\frac{\del}{\del \theta^{\pm}} 
- \I \ol{\theta}{}^{\pm} \big( \del_0 \pm \del_1 \big)
\, , &
\ol{D}{}_{\pm} \ &= \ 
- \frac{\del}{\del \ol{\theta}{}^{\pm}} 
+ \I \theta^{\pm} \big( \del_0 \pm \del_1 \big)
\, , \label{D_olD} \\
Q_{\pm} \ &= \ 
\frac{\del}{\del \theta^{\pm}} 
+ \I \ol{\theta}{}^{\pm} \big( \del_0 \pm \del_1 \big)
\, , &
\ol{Q}{}_{\pm} \ &= \ 
- \frac{\del}{\del \ol{\theta}{}^{\pm}} 
- \I \theta^{\pm} \big( \del_0 \pm \del_1 \big)
\, . \label{Q_olQ}
\end{align}
\esubeq
It is convenient to define integral measures of the Grassmann coordinates in superspace,
\bsubeq \label{f-measure-22}
\begin{gather}
\d^2 \theta \ \equiv \ 
- \frac{1}{4} \, \d \theta^{\alpha} \, \d \theta^{\beta} \,
\ve_{\alpha \beta} 
\ = \ 
- \half \, \d \theta^+ \, \d \theta^- 
\, , \ls
\d^2 \ol{\theta} \ \equiv \ 
- \frac{1}{4} \, \d \ol{\theta}{}_{\dot{\alpha}} \, 
\d \ol{\theta}{}_{\dot{\beta}} \, \ve^{\dot{\alpha} \dot{\beta}} 
\ = \ 
\half \, \d \ol{\theta}{}^+ \, \d \ol{\theta}{}^- 
\, , \\
\d^2 \wt{\theta} \ \equiv \ 
- \half \, \d \theta^+ \, \d \ol{\theta}{}^- 
\, , \ls 
\d^2 \ol{\wt{\theta}} \ \equiv \ 
- \half \, \d \theta^- \, \d \ol{\theta}{}^+ 
\, , \\
\d^4 \theta \ = \ \d^2 \theta \, \d^2 \ol{\theta} 
\ = \ 
- \d^2 \wt{\theta} \, \d^2 \ol{\wt{\theta}} 
\ = \ - \frac{1}{4} \d \theta^+ \, \d \theta^- \, \d \ol{\theta}{}^+ \,
\d \ol{\theta}{}^- 
\, , \\
\int \! \d^2 \theta \, \theta \theta \ = \ 1 
\, , \ls
\int \! \d^2 \ol{\theta} \, \ol{\theta} \ol{\theta} \ = \ 1 
\, , \ls
\int \! \d^2 \wt{\theta} \, \theta^+ \ol{\theta}{}^- \ = \ \half 
\, , \ls
\int \! \d^2 \ol{\wt{\theta}} \, \theta^- \ol{\theta}{}^+ \ = \ \half
\, .
\end{gather}
\esubeq
We note that hermitian conjugate of anti-commuting fermions is defined as $(\eta_+ \lambda_-)^{\dagger} = + \ol{\lambda}{}_- \ol{\eta}{}_+$.

\subsection{String worldsheet sigma model}

We define the normalization of string worldsheet sigma model.
This is important for discussing T-duality on the target space configuration.
Action and Lagrangian is described as
\bsubeq \label{string-NLSM-bosons}
\begin{gather}
S \ = \ 
\frac{1}{2 \pi \alpha'} \int \d^2 x \sqrt{- g} \, \Scr{L}
\, , \\
\Scr{L}
\ = \ 
- \half G_{MN} \, g^{mn} \, \del_m X^M \del_n X^N
+ \half B_{MN} \, \ve^{mn} \, \del_m X^M \del_n X^N
\, . 
\end{gather}
\esubeq
Here $X^M$ is a scalar field which represents the target space coordinate. 
and $G_{MN}$ and $B_{MN}$ denote the target space metric and B-field respectively.
They should follow the equations of motion in supergravity.
The target space dilaton does not appear in this sigma model 
if the worldsheet metric $g_{mn}$ is flat.
We set the normalization of the Levi-Civita invariant tensor $\ve^{mn}$ on the flat space $\ve^{01} = +1 = \ve^{10}$.
%
We exhibit Table \ref{ST-WS} with the assignment between the worldsheet scalar fields and the spacetime coordinates.
\begin{table}[h]
\begin{center}
\slb{.95}{\renewcommand{\arraystretch}{1.4}
\begin{tabular}{r|cccccc:cccc}
\hline
spacetime coordinates & 0 & 1 & 2 & 3 & 4 & 5 & 6 & 7 & 8 & 9 
\\ \hline 
NS5-brane &
$\bigcirc$ & $\bigcirc$ & $\bigcirc$ & $\bigcirc$ & $\bigcirc$ & $\bigcirc$ & 
$r^1 = x^6$ & $r^3 = x^7$ & $r^2 = x^8$ & $r^4 = x^9$
\\ 
KK5-brane & 
$\bigcirc$ & $\bigcirc$ & $\bigcirc$ & $\bigcirc$ & $\bigcirc$ & $\bigcirc$ & 
$r^1 = x^6$ & $r^3 = x^7$ & $r^2 = x^8$ & $\wt{r}^4 = \wt{x}^9$
\\ 
exotic $5^2_2$-brane &
$\bigcirc$ & $\bigcirc$ & $\bigcirc$ & $\bigcirc$ & $\bigcirc$ & $\bigcirc$ & 
$r^1 = x^6$ & $r^3 = x^7$ & $y^2 = \wt{x}^8$ & $\wt{r}^4 = \wt{x}^9$
\\ \hline
\end{tabular}
}
\caption{Assignment between the worldsheet fields and the spacetime coordinates.}
\label{ST-WS}
\end{center}
\end{table}

\section{Descriptions in supergravity}
\label{app:SUGRA}

In this appendix we briefly summarize the feature of five-branes in the supergravity framework. 

\subsection{Buscher rule}

We exhibit the Buscher rule \cite{Buscher:1987sk}, i.e., the T-duality transformation rule in supergravity.
When we perform T-duality along the $n$-th direction,
the spacetime metric $G_{MN}$, B-field $B_{MN}$ and dilaton $\phi$ are transformed as
\bsubeq \label{Buscher}
\begin{alignat}{2}
G'_{MN} \ &= \ 
G_{MN} - \frac{G_{nM} G_{nN} - B_{nM} B_{nN}}{G_{nn}}
\, , &\ls
G'_{nN} \ &= \ 
\frac{B_{bN}}{G_{nn}}
\, , \ls
G'_{nn} \ = \ 
\frac{1}{G_{nn}}
\, , \\
B'_{MN} \ &= \ 
B_{MN} + \frac{G_{nM} B_{Nn} - G_{nN} B_{Mn}}{G_{nn}}
\, , &\ls
B'_{nN} \ &= \ 
\frac{G_{nN}}{G_{nn}}
\, , \\
\phi' \ &= \ 
\phi - \half \log (G_{nn})
\end{alignat}
\esubeq
In the main part of this article we often use this rule.
The explicit form is necessary for avoiding any sign ambiguities from involution of B-field into duality transformations.

\subsection{Standard five-branes}

We begin with an NS5-brane smeared along one transverse direction.
This is also referred to as an H-monopole.
The background configuration in the string frame is given by 
\bsubeq \label{H-monopole}
\begin{gather}
\d s^2 \ = \ 
\d s^2_{012345} 
+ H \Big\{ (\d x^6)^2 + (\d x^7)^2 + (\d x^8)^2 + (\d x^9)^2 \Big\}
\, , \\
B_{i9} \ = \ 
\Omega_i
\, , \ls
\e^{2 \phi} \ = \ H
\, , \\
H \ = \ 
1 + \frac{\ell_0}{\sqrt{2} |\vec{x}|}
\, , \ls
\ell_0 \ = \ \frac{\alpha'}{{\cal R}_9}
\, , \\
\nabla_i H \ = \ 
(\nabla \times \vec{\Omega})_i
\, , \ls
\vec{\Omega} \cdot \d \vec{x} 
\ = \ 
\frac{\ell_0}{\sqrt{2}} \frac{- x^6 \, \d x^8 + x^8 \, \d x^6}{|\vec{x}|(|\vec{x}| + x^7)}
\, . 
\end{gather}
\esubeq
Here $\alpha'$ is the Regge parameter in string theory.
The NS5-brane is expanded in the 012345-directions whose spacetime metric is flat, while the transverse space of the 6789-direction is ${\mathbb R}^3 \times S^1$.
The vector $\vec{x}$ lives in the transverse 678-directions ${\mathbb R}^3$.
This five-brane is smeared along the transverse 9-th compact direction whose radius is ${\cal R}_9$.
The configuration (\ref{H-monopole}) is governed by a harmonic function $H$.
The B-field is given by a function $\Omega_i$ which is subject to the monopole equation, where the index $i$ runs $i = 6,7,8$.


Next we mention a KK-monopole, or referred to as a KK5-brane.
This is obtained via T-duality along the 9-th direction of the H-monopole (\ref{H-monopole}),
\bsubeq \label{KK-monopole}
\begin{gather}
\d s^2 \ = \ 
\d s^2_{012345} 
+ H \Big\{ (\d x^6)^2 + (\d x^7)^2 + (\d x^8)^2 \Big\}
+ \frac{1}{H} \Big( \d \wt{x}^9 - \vec{\Omega} \cdot \d \vec{x} \Big)^2
\, , \\
B_{MN} \ = \ 0
\, , \ls
\e^{2 \phi} \ = \ 1
\, .
\end{gather}
\esubeq
Due to T-duality, the B-field in the H-monopole configuration is involved into the off-diagonal part of the spacetime metric as the KK-vector $\vec{\Omega}$.
We also see that the dilaton becomes trivial.
The transverse space in the 6789-directions becomes the Taub-NUT space, a noncompact hyper-K\"{a}hler geometry.
In order to emphasize the T-duality transformation along the 9-th direction, 
we refer to this coordinate as $\wt{x}^9$ whose radius is $\wt{\cal R}_9$.

\subsection{Defect five-branes}

In the previous subsection we mentioned two standard five-branes.
They are of codimension three.
Here we consider five-branes of codimension two, i.e., defect five-branes \cite{Bergshoeff:2011se}.
We can easily find defect five-branes from the H-monopole and the KK5-brane if one of the transverse direction is further smeared\footnote{The smearing procedure is discussed in \cite{deBoer:2010ud, Kikuchi:2012za, Kimura:2013fda} and so forth.}.
One of the most interesting defect five-branes is an exotic $5^2_2$-brane.


We first discuss a defect NS5-brane smeared along the 8-th direction of the H-monopole (\ref{H-monopole}).
The background configuration is given as
\bsubeq \label{dNS5}
\begin{gather}
\d s^2 \ = \ 
\d s^2_{012345} 
+ H_{\sigma} \Big\{ (\d \varrho)^2 + \varrho^2 (\d \vartheta)^2 \Big\}
+ H_{\sigma} \Big\{ (\d x^8)^2 + (\d x^9)^2 \Big\}
\, , \\
B_{89} \ = \ \Omega_{\sigma}
\, , \ls
\e^{2 \phi} \ = \ H_{\sigma}
\, , \\
H_{\sigma} \ = \ 
h + \sigma \, \log \frac{\mu}{\varrho}
\, , \ls
\Omega_{\sigma} \ = \ \sigma \, \vartheta
\, , \\
x^6 \ = \ \varrho \cos \vartheta
\, , \ls
x^7 \ = \ \varrho \sin \vartheta
\, , \ls
\sigma \ = \ \frac{\ell_0}{2 \pi {\cal R}_8}
\, .
\end{gather}
\esubeq
Here ${\cal R}_8$ is the radius of the compact circle along the smeared 8-th direction.
Now the space in the 89-directions becomes a two-torus $T^2_{89}$.
We notice that the harmonic function $H$ is reduced to a logarithmic function $H_{\sigma}$,
where $\mu$ is the renormalization scale and $h$ is the bare quantity which diverges if we go infinitely far away from the five-brane.
In this sense the representation (\ref{dNS5}) is valid only close to the defect NS5-brane.
The B-field seems ill-defined under shift of the angular coordinate $\vartheta \to \vartheta + 2\pi$. 
However, we can always remove this feature by the B-field gauge transformation.


There exist two isometries along the 8-th and 9-th directions of the defect NS5-brane (\ref{dNS5}).
Taking T-duality along the 9-th direction $x^9$ to $\wt{x}^9$,
we obtain a defect KK5-brane (or called a KK-vortex \cite{Okada:2014wma}),
\bsubeq \label{dKK5}
\begin{gather}
\d s^2 \ = \ 
\d s^2_{012345}
+ H_{\sigma} \Big\{ (\d \varrho)^2 + \varrho^2 (\d \vartheta)^2 \Big\}
+ H_{\sigma} (\d x^8)^2
+ \frac{1}{H_{\sigma}} \Big( \d \wt{x}^9 - \Omega_{\sigma} \, \d x^8 \Big)^2
\, , \\
B_{MN} \ = \ 0
\, , \ls
\e^{2 \phi} \ = \ 1
\, . 
\end{gather}
\esubeq
This is also found if the KK5-brane (\ref{KK-monopole}) is smeared along the 8-th direction.
The space in the 89-direction is a two-torus $T^2_{89}$ twisted by the KK-vector $\Omega_{\sigma}$.
Here the B-field and dilaton are again trivial.
The spacetime metric on $T^2_{89}$ seems ill-defined under shift $\vartheta \to \vartheta + 2\pi$. 
However, since this can be removed by the coordinate transformation,
we can understand that the spacetime metric is single-valued.
Actually this space is regarded as an ALG space, a generalization of an ALF space which asymptotically has a tri-holomorphic two-torus action \cite{Cherkis:2000cj, Cherkis:2001gm}.

If we take T-duality along the 8-th direction instead of the 9-th direction of the defect NS5-brane (\ref{dNS5}),
we find another defect KK5-brane (or referred to as an anti KK-vortex \cite{Okada:2014wma}).
Its background configuration is given as
\bsubeq \label{AK5}
\begin{gather}
\d s^2 \ = \ 
\d s^2_{012345} 
+ H_{\sigma} \Big\{ (\d \varrho)^2 + \varrho^2 (\d \vartheta)^2 \Big\}
+ H_{\sigma} (\d x^9)^2
+ \frac{1}{H_{\sigma}} \Big( \d \wt{x}^8 + \Omega_{\sigma} \, \d x^9 \Big)^2
\, , \\
B_{MN} \ = \ 0
\, , \ls
\e^{2 \phi} \ = \ 1
\, . 
\end{gather}
\esubeq
Here $\wt{x}^8$ is the dual coordinate of  $x^8$.
This object plays a significant role in the analysis of Alice string \cite{Okada:2014wma}.


Performing T-duality along the 8-th direction of the defect KK5-brane (\ref{dKK5}) (or the 9-th direction of (\ref{AK5})),
we obtain an exotic $5^2_2$-brane \cite{deBoer:2010ud, Kikuchi:2012za, deBoer:2012ma, Kimura:2013fda} whose background configuration is
\bsubeq \label{522}
\begin{gather}
\d s^2 \ = \ 
\d s^2_{012345} 
+ H_{\sigma} \Big\{ (\d \varrho)^2 + \varrho^2 (\d \vartheta)^2 \Big\}
+ \frac{H_{\sigma}}{K_{\sigma}} \Big\{ (\d \wt{x}^8)^2 + (\d \wt{x}^9)^2 \Big\}
\, , \\
B_{89} \ = \ 
- \frac{\Omega_{\sigma}}{K_{\sigma}}
\, , \ls
\e^{2 \phi} \ = \ 
\frac{H_{\sigma}}{K_{\sigma}}
\, , \ls
K_{\sigma} \ = \ (H_{\sigma})^2 + (\Omega_{\sigma})^2
\, .
\end{gather}
\esubeq
The space in the 89-directions is again a two-torus $T^2_{89}$.
Here the B-field and dilaton are non-trivial as in the configuration of the defect NS5-brane (\ref{dNS5}).
Indeed, not only the spacetime metric, but also the B-field and dilaton are no longer single-valued under shift of the angular coordinate $\vartheta \to \vartheta + 2 \pi$.
It is impossible to remove this shift by the coordinate transformations or the B-field gauge transformation.
This feature originates from the T-duality structure $O(2,2;{\mathbb Z})$ on the two-torus $T^2_{89}$.
This is the reason why this five-brane is ``exotic''.

\end{appendix}

}

\begin{thebibliography}{99}

\bibitem{Kimura:2013fda}
  T.~Kimura and S.~Sasaki,
  ``{\sl Gauged linear sigma model for exotic five-brane},''
  Nucl.\ Phys.\ B {\bf 876} (2013) 493
  [arXiv:1304.4061 [hep-th]].

\bibitem{Polchinski:1995mt}
  J.~Polchinski,
  ``{\sl Dirichlet branes and Ramond-Ramond charges},''
  Phys.\ Rev.\ Lett.\  {\bf 75} (1995) 4724
  [hep-th/9510017].

\bibitem{Strominger:1990et}
  A.~Strominger,
  ``{\sl Heterotic solitons},''
  Nucl.\ Phys.\ B {\bf 343} (1990) 167
   [Erratum-ibid.\ B {\bf 353} (1991) 565].

\bibitem{Sorkin:1983ns}
  R.~D.~Sorkin,
  ``{\sl Kaluza-Klein monopole},''
  Phys.\ Rev.\ Lett.\  {\bf 51} (1983) 87.

\bibitem{Blau:1997du}
  M.~Blau and M.~O'Loughlin,
  ``{\sl Aspects of U-duality in matrix theory},''
  Nucl.\ Phys.\ B {\bf 525} (1998) 182
  [hep-th/9712047].

\bibitem{Obers:1998fb}
  N.~A.~Obers and B.~Pioline,
  ``{\sl U-duality and M-theory},''
  Phys.\ Rept.\  {\bf 318} (1999) 113
  [hep-th/9809039].

\bibitem{Bergshoeff:2011se}
  E.~A.~Bergshoeff, T.~Ort\'{\i}n and F.~Riccioni,
  ``{\sl Defect branes},''
  Nucl.\ Phys.\ B {\bf 856} (2012) 210
  [arXiv:1109.4484 [hep-th]].

\bibitem{Witten:1993yc}
  E.~Witten,
  ``{\sl Phases of $\N=2$ theories in two dimensions},''
  Nucl.\ Phys.\ B {\bf 403} (1993) 159
  [hep-th/9301042].

\bibitem{Hori:2000kt}
  K.~Hori and C.~Vafa,
  ``{\sl Mirror symmetry},''
  hep-th/0002222.

\bibitem{Douglas:1996sw}
  M.~R.~Douglas and G.~W.~Moore,
  ``{\sl D-branes, quivers, and ALE instantons},''
  hep-th/9603167.

\bibitem{Johnson:1996py}
  C.~V.~Johnson and R.~C.~Myers,
  ``{\sl Aspects of type IIB theory on ALE spaces},''
  Phys.\ Rev.\ D {\bf 55} (1997) 6382
  [hep-th/9610140].

\bibitem{Giveon:1998sr}
  A.~Giveon and D.~Kutasov,
  ``{\sl Brane dynamics and gauge theory},''
  Rev.\ Mod.\ Phys.\  {\bf 71} (1999) 983
  [hep-th/9802067].

\bibitem{Tong:2002rq}
  D.~Tong,
  ``{\sl NS5-branes, T-duality and worldsheet instantons},''
  JHEP {\bf 0207} (2002) 013
  [hep-th/0204186].

\bibitem{Harvey:2005ab}
  J.~A.~Harvey and S.~Jensen,
  ``{\sl Worldsheet instanton corrections to the Kaluza-Klein monopole},''
  JHEP {\bf 0510} (2005) 028
  [hep-th/0507204].

\bibitem{Okuyama:2005gx}
  K.~Okuyama,
  ``{\sl Linear sigma models of H and KK monopoles},''
  JHEP {\bf 0508} (2005) 089
  [hep-th/0508097].

\bibitem{deBoer:2010ud} 
  J.~de Boer and M.~Shigemori,
  ``{\sl Exotic branes and non-geometric backgrounds},''
  Phys.\ Rev.\ Lett.\  {\bf 104} (2010) 251603
  [arXiv:1004.2521 [hep-th]].

\bibitem{Kikuchi:2012za} 
  T.~Kikuchi, T.~Okada and Y.~Sakatani,
  ``{\sl Rotating string in doubled geometry with generalized isometries},''
  Phys.\ Rev.\ D {\bf 86} (2012) 046001
  [arXiv:1205.5549 [hep-th]].

\bibitem{deBoer:2012ma}
  J.~de Boer and M.~Shigemori,
  ``{\sl Exotic branes in string theory},''
  Phys.\ Rept.\  {\bf 532} (2013) 65
  [arXiv:1209.6056 [hep-th]].

\bibitem{Rocek:1991ps}
  M.~Ro\v{c}ek and E.~P.~Verlinde,
  ``{\sl Duality, quotients, and currents},''
  Nucl.\ Phys.\ B {\bf 373} (1992) 630
  [hep-th/9110053].

\bibitem{Kimura:2014bxa}
  T.~Kimura and M.~Yata,
  ``{\sl Gauged linear sigma model with F-term for A-type ALE space},''
  PTEP {\bf 2014} (2014) 7,  073B01
  [arXiv:1402.5580 [hep-th]].

\bibitem{Kimura:2014aja}
  T.~Kimura and M.~Yata,
  ``{\sl T-duality transformation of gauged linear sigma model with F-term},''
  Nucl.\ Phys.\ B {\bf 887} (2014) 136
  [arXiv:1406.0087 [hep-th]].

\bibitem{Kimura:2013zva}
  T.~Kimura and S.~Sasaki,
  ``{\sl Worldsheet instanton corrections to $5^2_2$-brane geometry},''
  JHEP {\bf 1308} (2013) 126
  [arXiv:1305.4439 [hep-th]].

\bibitem{Kimura:2013khz}
  T.~Kimura and S.~Sasaki,
  ``{\sl Worldsheet description of exotic five-brane with two gauged isometries},''
  JHEP {\bf 1403} (2014) 128
  [arXiv:1310.6163 [hep-th]].

\bibitem{Harvey:2014nha}
  J.~A.~Harvey, S.~Lee and S.~Murthy,
  ``{\sl Elliptic genera of ALE and ALF manifolds from gauged linear sigma models},''
  JHEP {\bf 1502} (2015) 110
  [arXiv:1406.6342 [hep-th]].

\bibitem{Sen:1997zb}
  A.~Sen,
  ``{\sl Kaluza-Klein dyons in string theory},''
  Phys.\ Rev.\ Lett.\  {\bf 79} (1997) 1619
  [hep-th/9705212].

\bibitem{Gregory:1997te}
  R.~Gregory, J.~A.~Harvey and G.~W.~Moore,
  ``{\sl Unwinding strings and T-duality of Kaluza-Klein and H monopoles},''
  Adv.\ Theor.\ Math.\ Phys.\  {\bf 1} (1997) 283
  [hep-th/9708086].

\bibitem{Berman:2014jsa}
  D.~S.~Berman and F.~J.~Rudolph,
  ``{\sl Branes are waves and monopoles},''
  arXiv:1409.6314 [hep-th].

\bibitem{Okada:2014wma}
 T.~Okada and Y.~Sakatani,
  ``{\sl Defect branes as Alice strings},''
  JHEP {\bf 1503} (2015) 131
  [arXiv:1411.1043 [hep-th]].

\bibitem{Hull:2004in}
  C.~M.~Hull,
  ``{\sl A geometry for non-geometric string backgrounds},''
  JHEP {\bf 0510} (2005) 065
  [hep-th/0406102].

\bibitem{Hull:2006va}
  C.~M.~Hull,
  ``{\sl Doubled geometry and T-folds},''
  JHEP {\bf 0707} (2007) 080
  [hep-th/0605149].

\bibitem{Dall'Agata:2008qz}
  G.~Dall'Agata and N.~Prezas,
  ``{\sl Worldsheet theories for non-geometric string backgrounds},''
  JHEP {\bf 0808} (2008) 088
  [arXiv:0806.2003 [hep-th]].

\bibitem{Andriot:2013xca}
  D.~Andriot and A.~Betz,
  ``{\sl $\beta$-supergravity: a ten-dimensional theory with non-geometric fluxes, and its geometric framework},''
  JHEP {\bf 1312} (2013) 083
  [arXiv:1306.4381 [hep-th]].

\bibitem{Andriot:2014uda}
  D.~Andriot and A.~Betz,
  ``{\sl NS-branes, source corrected Bianchi identities, and more on backgrounds with non-geometric fluxes},''
  JHEP {\bf 1407} (2014) 059
  [arXiv:1402.5972 [hep-th]].

\bibitem{Sakatani:2014hba}
  Y.~Sakatani,
  ``{\sl Exotic branes and non-geometric fluxes},''
  JHEP {\bf 1503} (2015) 135
  [arXiv:1412.8769 [hep-th]].


\bibitem{Ooguri:1996me}
  H.~Ooguri and C.~Vafa,
  ``{\sl Summing up D instantons},''
  Phys.\ Rev.\ Lett.\  {\bf 77} (1996) 3296
  [hep-th/9608079].

\bibitem{Chatzistavrakidis:2013jqa}
  A.~Chatzistavrakidis, F.~F.~Gautason, G.~Moutsopoulos and M.~Zagermann,
  ``{\sl Effective actions of nongeometric five-branes},''
  Phys.\ Rev.\ D {\bf 89} (2014) 6,  066004
  [arXiv:1309.2653 [hep-th]].

\bibitem{Kimura:2014upa}
  T.~Kimura, S.~Sasaki and M.~Yata,
  ``{\sl World-volume effective actions of exotic five-branes},''
  JHEP {\bf 1407} (2014) 127
  [arXiv:1404.5442 [hep-th]].

\bibitem{Bergshoeff:1997gy}
  E.~Bergshoeff, B.~Janssen and T.~Ort\'{\i}n,
  ``{\sl Kaluza-Klein monopoles and gauged sigma models},''
  Phys.\ Lett.\ B {\bf 410} (1997) 131
  [hep-th/9706117].

\bibitem{Kimura:2014wga}
  T.~Kimura,
  ``{\sl Defect $(p,q)$ five-branes},''
  Nucl.\ Phys.\ B {\bf 893} (2015) 1
  [arXiv:1410.8403 [hep-th]].

\bibitem{Kimura:2014bea}
  T.~Kimura, S.~Sasaki and M.~Yata,
  ``{\sl Hyper-K\"{a}hler with torsion, T-duality, and defect $(p,q)$ five-branes},''
  JHEP {\bf 1503} (2015) 076
  [arXiv:1411.3457 [hep-th]].

\bibitem{Hellerman:2002ax}
  S.~Hellerman, J.~McGreevy and B.~Williams,
  ``{\sl Geometric constructions of nongeometric string theories},''
  JHEP {\bf 0401} (2004) 024
  [hep-th/0208174].

\bibitem{Greene:1989ya}
  B.~R.~Greene, A.~D.~Shapere, C.~Vafa and S.~T.~Yau,
  ``{\sl Stringy cosmic strings and noncompact Calabi-Yau manifolds},''
  Nucl.\ Phys.\ B {\bf 337} (1990) 1.

\bibitem{Bergshoeff:2006jj}
  E.~A.~Bergshoeff, J.~Hartong, T.~Ort\'{\i}n and D.~Roest,
  ``{\sl Seven-branes and supersymmetry},''
  JHEP {\bf 0702} (2007) 003
  [hep-th/0612072].

\bibitem{Buscher:1987sk}
  T.~H.~Buscher,
  ``{\sl A symmetry of the string background field equations},''
  Phys.\ Lett.\ B {\bf 194} (1987) 59.

\bibitem{Cherkis:2000cj}
  S.~A.~Cherkis and A.~Kapustin,
  ``{\sl Nahm transform for periodic monopoles and $\N=2$ super Yang-Mills theory},''
  Commun.\ Math.\ Phys.\  {\bf 218} (2001) 333
  [hep-th/0006050].

\bibitem{Cherkis:2001gm}
  S.~A.~Cherkis and A.~Kapustin,
  ``{\sl Hyperk\"{a}hler metrics from periodic monopoles},''
  Phys.\ Rev.\ D {\bf 65} (2002) 084015
  [hep-th/0109141].


\end{thebibliography}
\end{document}